\documentclass[twocolumn,a4]{revtex4-1}
\usepackage{relsize}
\usepackage[countmax]{subfloat}
\usepackage{graphicx}
\usepackage{amssymb, amsmath,amssymb,amsfonts}
\usepackage{amsthm,mathrsfs,amsopn}
\usepackage{dcolumn}
\usepackage{bm}
\usepackage{color}
\usepackage[utf8]{inputenc}
\usepackage[table]{xcolor} 
\usepackage{floatrow}
\usepackage{lettrine}
\usepackage{subfigure}
\usepackage{dsfont}
\usepackage{hyperref}
\hypersetup{
    colorlinks=true,
    linkcolor=blue,
    filecolor=magenta,      
    urlcolor=cyan,
    citecolor = blue,
}

\definecolor{BE}{RGB}{13,55,13}
\begin{document}

\title{The role of modularity in self-organisation dynamics in biological networks}

\author{Bram A. Siebert$^{1}$, Cameron L. Hall$^{1,2}$, James P. Gleeson$^{1}$, Malbor Asllani$^{1}$ \vspace*{.25cm}}
\affiliation{$^1$MACSI, Department of Mathematics and Statistics, University of Limerick, Limerick V94 T9PX, Ireland}
\affiliation{$^2$Department of Engineering Mathematics, University of Bristol, Bristol Woodland Road, Clifton BS8 1UB, UK}

\begin{abstract}

Interconnected ensembles of biological entities are perhaps some of the most complex systems that modern science has encountered so far. 
In particular, scientists have concentrated on understanding how the complexity of the interacting structure between different neurons, proteins or species influences the functioning of their respective systems. 
It is well-established that many biological networks are constructed in a highly hierarchical way with two main properties: short average paths that join two apparently distant nodes (neuronal, species, or protein patches) and a high proportion of nodes in modular aggregations.
Although several hypotheses have been proposed so far, still little is known about the relation of the modules with the dynamical activity in such biological systems. 
Here we show that network modularity is a key ingredient for the formation of self-organising patterns of functional activity, independently of the topological peculiarities of the structure of the modules. 
In particular,
we propose 
%the first 
a self-organising mechanism which explains the formation of macroscopic spatial patterns, which are homogeneous within modules. This may explain how spontaneous order in biological networks follows their modular structural organisation. %Our results explain the relation between the modular structure and the modules as functional units in biological networks. The latter result asserts that 
We test our results on real-world networks to confirm the important role of modularity in creating macro-scale patterns. 
\end{abstract}

\maketitle

\section{Introduction}
\label{sec:intro}

Patterns are macroscopic structures that are the distinctive mark of the self-organisation in a system of microscopic interacting entities \cite{Nicolis1977}. 
They are ubiquitous in nature and can be seen in the spots of a leopard’s fur or the coloured scales of a butterfly’s wing \cite{Murray2001}. 
In 1952, Alan Turing published his seminal work on pattern formation, \textit{The Chemical Basis of Morphogenesis} where he laid down {an elegant and plausible theory that can be used to explain} the formation of patterns \cite{Turing1952}. 
Turing developed a simple model of pattern formation that established the minimal requirements for a biochemical system to self-organise. 
Turing's minimal system is composed of two ``competing'' chemicals, an activator and an inhibitor, which share the same spatial domain where they react and diffuse.
Based on a diffusion-driven instability mechanism, today known as Turing instability, Turing showed that it is possible to explain {and predict the growth of spatially inhomogeneous perturbations away from a spatially homogeneous steady state. 
These perturbations in concentration} are later stabilised by nonlinearities in the system, yielding the celebrated Turing patterns. 
%\textcolor{magenta}{I don't like ``it has been shown''. By whom has it been shown? Additionally, what do we mean by short-range activation and long-range inhibition? Is the important thing that inhibitor needs to diffuse faster? I would initially read short-range and long-range as implying non-local interactions between species, but I know that's not right.} 
It can be shown that the right combination of short-range activation and long-range inhibition{, caused by slowly diffusing activators and rapidly diffusing inhibitors,} enables the pattern forming phenomenon~\cite{Gierer1972}. 

{Conventionally, an activator-inhibitor system is modelled using a set of reaction-diffusion equations that describe the evolution of the concentrations of activator and inhibitor throughout a continuous medium. These equations can readily be adapted to describe activator-inhibitor systems in discrete systems such as regular lattices, and they have been used in this way to describe pattern formations in cellular tissues \cite{Othmer1971, Gierer1972}.}
However, biological tissue often takes more complex forms, and the spatial support cannot always be adequately formulated via regular lattices. %{\bf[JG: Not clear, can you rephrase? BS:Done]}. 
%represented causing the dynamics to skip a standard formulation through PDEs 
Inspired by the network structures of early stages of embryogenesis \cite{Schnabel2006}, ecological meta-populations \cite{Holland2008} or coupled chemical reactors \cite{Horsthemke2004}, researchers have extended the reaction-diffusion formalism to complex biological networks~\cite{Othmer1971, Nakao2010, Asllani2014Multiplex, Asllani2014Directed, Asllani2015, Asllani2016, riccardo, AsllaniEPJB20}. 
These discrete structures {consist of} graphs where the nodes usually represent the cells inside which reactions occur, and the edges usually represent the routes through which cells communicate by exchanging chemicals. 

%\textcolor{magenta}{Avoid passive voice for references where possible! H{\"u}tt \emph{et al.}\ \cite{Hutt} have recently argued... }
H{\"u}tt et al. \cite{Hutt} recently argued that the formalism of activator-inhibitor systems is relevant to  the dynamical processes evolving in the brain \cite{Hutt} .
%H{\"u}tt \emph{et al.}\ \cite{Hutt} recently argued that this formalism \textcolor{green}{which formalism? perhaps say ``activator-inhibitor systems on biological networks''? And even then, it's not immediately clear how a formalism influences an interpretation? I'm not sure what is being meant here.} also influences the interpretation of the dynamical processes evolving in the brain \cite{Hutt}.
The implementation of network tools for analysing the brain's structure has been used since the first years of network science \cite{Watts1998}. 
%Using sophisticated techniques such as the functional Magnetic Resonance Imaging (fMRI) neuroscientists discovered that they are two categories of brain networks: the structural ones constituted by neurons or neural patches connected together through axons and functional networks known also as connectomes. In the latter large group of neurons characterised by very similar dynamical behaviour makes up the individual nodes which are linked together through functional routes according to correlation protocols in the dynamical response between nodes.
%\textcolor{magenta}{The following sentence doesn't make sense after ``discovered that''.} 
In their seminal work, Watts and Strogatz \cite{Watts1998} studied the topology of the neuronal network of the nematode \textit{C.\ elegans} and discovered that these networks possess a ``small-world'' property. 
In the literature, it has also been argued that many brain networks might be small-world networks \cite{Meunier2010,HarrigerLogan2012,Hahn2019}. 
It is widely accepted that the small-world property of brain connectomes should help the communication between neurons inside the brain by integrating multiple segregated sources of information \cite{SpornsBook}.        

A further property of brain networks is that they are often modular \cite{Meunier2010} so that the neurons can be segregated into communities (referred to as modules) where two neurons chosen at random from the same module are much more likely to be connected than two neurons chosen at random from different modules. 
The functional role that the modularity of brain connections has been discussed from several perspectives. 
For example, due to the increased structural stability~\cite{Simon1962,SpornsBook}, the modularity might have been crucial in the evolution and development of the brain. 
%\textcolor{magenta}{What wiring cost and how optimised? Is there a citation possible here? Perhaps ``In the case of a spatial network where the cost of a connection is an increasing function of the distance between nodes, modular topology can also be used to reduce the wiring cost.'' But it is not clear to me why the wiring cost should not simply be minimised by a carefully-constructed tree (e.g. from a greedy algorithm adding shortest connections first). I don't quite get what we are getting at here at all.}
According to \cite{Meunier2010,Simon1962} modular topology can also optimise the wiring cost in the case of spatial networks. 
A small number of long range (and thus costly) connections reduces the diameter of the network, and allows the remaining nodes, now grouped into communities or modules, to form dense small world networks.
Also, more compact segregation of neurons may contribute to the specialisation of the neurons in their functional duties~\cite{SpornsBook}. 
To ensure both a low shortest path length, and a high clustering coefficient, brain networks are organised in a strict hierarchical manner~\cite{Sporns2007, Bullmore2009, Meunier2010} where at the first level of the hierarchy sets of nodes (the modules) are connected to mimic a small-world topology and the same happens at the second level of hierarchy and so on, until the single node level. For a more detailed discussion of the role of the hierarchy in the pattern formation process see the Appendix.

More generally, modularity is a common topological property that naturally emerges in biological, ecological, and social scenarios where the different communities are associated with different functions of the system represented by the network as a whole \cite{NewmanPNAS}. 
There are many examples of this:
in protein interaction networks, the proteins that share  similar functions are grouped together in modules \cite{protein}; in metabolic networks, there are structural/functional communities corresponding to cycles or pathways \cite{metabolic}; and in citation networks, scientific papers are clustered according to their research topic \cite{Redner}.
%\textcolor{magenta}{The rest of the paragraph above has talked about the presence of modularity in various networks. Consequently, the following sentence sounds like we are proposing a mechanism by which modularity can arise? I think this needs to be rephrased to make it clear that we are looking at self-organisation (in the sense of pattern development) in systems where modularity is already present.}
In addition to these properties, in this paper, we propose a new mathematical mechanism that highlights the role that modularity {takes} in self-organising processes in biological networks.% (with a particular focus on models of brain networks) {\bf[JG: I think the claims about brain networks are too strong throughout. The results are certainly of potential interest for applicability to brain networks, and we can speculate in that direction in the Discussion section, but I don't think we can say that we work directly on brain networks here (and the models are all highly stylised too). I'd argue that we prove a concept here, and then argue (but not really show) that it may be important in brain and other biological networks]}.

Using the Turing theory of pattern formation, we show that spatially extended patterns can be triggered by the segregation of the nodes (neurons) in distinguishable communities. 
%\textcolor{magenta}{``known in the literature'' but no reference to the literature? In fairness, I feel that dispersion relations are key to linear stability analysis and don't really need an introduction. I would suggest something along the lines of ``We take a linear stability approach and show that modular networks can easily yield dispersion relations associated with pattern formation. As we will show, this is a consequence of modular networks (in contrast to other networks, e.g., small world ones) having a small spectral gap...''  Note also that we seem to be using spectral gap here in a different sense from how we use it in the caption of Figure 1 d). There may need to be clarity as to whether the spectral gap is between the first and second eigenvalues or between the last `modular' and first `non-modular' eigenvalues.}
To formally analyse the chances of such networks self-organising, we use a linear stability approach known in the literature as the dispersion relation \cite{Murray2001}.
We focus on modular networks, which in contrast to many other random networks, are characterised by a small spectral gap, i.e., a small distance of the second largest eigenvalue \footnote{The definition of the spectral gap depends on the way one defines the Laplacian matrix \cite{NewmanBook}. In our case the spectrum of the Laplacian is non positive.} of the Laplacian from the origin. Let us notice here that a small spectral gap is a characteristic also of large (dense) regular graphs, however, here we focus on random graphs.
To anticipate some of the technical details, we discuss the key features of modular networks in the following paragraphs and outline how these affect pattern formation.

For modular networks, the Laplacian eigenvalues that may be responsible for the Turing instability can be split into two sets.
In one set, we have the eigenvalues emerging due to the global modularity of the network, which we denote as ``modular eigenvalues''. 
In Sec.~\ref{sec:whymodular}, we will show that when only this part of the spectrum is {responsible for the} instability, then the shape of the associated pattern follows that of the network in the sense that nodes belonging to the same modules have very similar concentrations of the species among themselves but these concentrations are distinctly different from the concentrations in other modules. 
In contrast, if the instability is caused by the remaining set of  eigenvalues, which correspond to the local %communities
connectivity of nodes, here denoted as ``non-modular eigenvalues'', then all the nodes have (in principle) different concentrations making the pattern globally heterogeneous. 
%\textcolor{magenta}{Latter case? This sounds more like the former case.}
In this latter case, if the eigenvalues responsible for the instability are limited to the eigenvalues belonging to a single module, then the pattern will first emerge in that module. 
%\textcolor{magenta}{I don't see what the ``thus'' connects to here. Could we unpack this further? Also, what do we mean by functional units here? Is it an inherent property of the nodes (like proteins sharing similar functions, as used in the earlier example), or an emergent property of the pattern formation? It sounds like the term is being used to refer to the pattern formation, but this feels like the opposite of the earlier example.}

We aim to create a  bridge between the role of the structure in many biological networks with the dynamical activity therein. In particular, in our model, we explain how communities of biological entities (cells, individuals, etc.) can act as functional units in their corresponding biological systems. 
As a consequence, we argue that this approach can potentially be used in community detection methods \cite{Fortunato,Newman2004,NewmanBook} for networked biological systems where Turing patterns are known to exist.
However, it is important to note that this method partitions the network in a similar fashion to the Fiedler partitioning. Therefore, it is possible to underestimate the total number of communities. Additionally, using pattern formation for community detection does not distinguish between functional communities and structural communities.

In this paper we begin in Sec.~\ref{sec1} with a description of the mathematical background of Turing patterns. 
This will lead us into a discussion as to why modularity is critical to the formation of patterns in Sec.~\ref{subsec:DvDu}.
We describe the different types of patterns which form in Sec.~\ref{sec:whymodular}, and show how increasing the modularity helps in the formation of patterns.
Finally in Sec.~\ref{sec:realNetworks} we look for Turing patterns in some real world networks.

%\section{Results}
\section{Pattern formation on a networked system}
\label{sec1}

In a continuous domain, {the most simple Turing mechanism is given} in terms of reaction-diffusion equations that describe the evolution {through time and space} of the concentrations of two competing chemical species, called the activator (with concentration denoted $u(x,t)$) and the inhibitor (with concentration denoted $v(x,t)$) \cite{Turing1952,Murray2001}. 
In general, an activator increases production of both itself and the inhibitor.
The inhibitor, in turn, slows down the growth in activator.
When the spatial support is instead discrete, constituted by spatial patches (nodes) connected through communicating routes (links) the reaction-diffusion mechanism can be formulated using ODEs, instead of PDEs \cite{Othmer1971}.
In general, a two-species reaction-diffusion model on a network of $N$ nodes will take the form, 
\begin{equation}
\begin{aligned}
\frac{d u_i}{d t} &=& f(u_i,v_i) + D_u \sum_j {L}_{ij}u_j,\, \forall i=1,\dots, N \\
\frac{d v_i}{d t} &= & g(u_i,v_i) + D_v \sum_j {L}_{ij}v_j,\, \forall i=1,\dots, N,
\end{aligned}
\label{eq:turingPatternNetworks}
\end{equation}
where $u_i$ and $v_i$ represent the concentrations of activator and inhibitor respectively at node $i$, $f$ and $g$ are nonlinear functions that describe the net production rates of activator and inhibitor respectively, $D_u$ and $D_v$ are the diffusion coefficients of activator and inhibitor respectively, and $\mathbf{L}$ is the graph Laplacian operator. The entries $L_{ij}$ of the graph Laplacian are defined by $L_{ij} = A_{ij} - k_{i} \delta_{ij}$, where $\mathbf{A}$ is the adjacency matrix, $k_i$ is the degree of node $i$, $\delta$ is the Kronecker delta, and where we do not sum over repeated indices.
%This minimalistic model is responsible for describing the evolution dynamics of the activator and inhibitor species respectively denoted by $u_i$ and $v_i$ for each node $i$. 
In order to understand the development of spatial patterns, we analyse the linear stability of the system starting from a homogeneous steady state $(u^*,v^*)$ that is stable in the absence of diffusion. 
If the diffusion coefficients are nonzero and the ratio $\rho=D_v/D_u$ is large enough, the steady state $(u^*,v^*)$ becomes unstable and small random perturbations of the previous steady state will grow. This growth is exponential in the initial linear regime, and may then be stabilised by the nonlinear terms of the functions $f$ and $g$ so that the system reaches a stable but spatially inhomogeneous steady state.
Such a mechanism is responsible for the emergence of Turing patterns. 

%\textcolor{magenta}{I would like to remove the ``thus'' below (since it's not clear what this is immediately following from) and I would define $\delta \mathbf{x} = (\delta \mathbf{u},\delta\mathbf{v}) = (\textbf{u}-u^*,\textbf{v}-v^*)$ instead of defining $\mathbf{x}$, which is ambiguous. (By the way, it shouldn't make any difference for these variables, but it's good practice to use mathbf rather than textbf in equations.) It may also be good practice to define the Jacobian more clearly. Another issue is that the matrix $\mathbf{L}$ below is $2N$ by $2N$, whereas the original matrix $L$ was $N$ by $N$. This is related to the notational issues I discuss later on.} 
The linearised system in matrix form reads:
{\begin{equation}
\frac{d (\delta \textbf{x})}{d t} = \left(\hat{\textbf{J}} + \textbf{D} \hat{\textbf{L}}\right)\delta \textbf{x},
\label{eq:turingPatternsLinearisedNetwork}
\end{equation}
where $\delta\mathbf{x}=(\textbf{u}-u^*\mathds{1}_N,\textbf{v}-v^*\mathds{1}_N)$ is the perturbations vector of the activator $\textbf{u}$ and inhibitor $\textbf{v}$ species, $\mathds{1}_N$ is the all-ones $N-$dimensional  vector, and 
\begin{equation*}
    \textbf{D} = 
    \begin{bmatrix}
    D_u\textbf{I}_N & 0 \\
    0 & D_v\mathbf{I}_N
    \end{bmatrix}
\end{equation*}
is the diffusion constant matrix.
Note that $\mathbf{I}_N$ represents the $N$ by $N$ identity matrix, so that $\mathbf{D}$ is $2N$ by $2N$.
The Jacobian matrix and the extended Laplacian are correspondingly 
\begin{equation*}
\hat{\textbf{J}}=
\begin{bmatrix}
f_u\textbf{I}_N  & f_v\textbf{I}_N \\
g_u\textbf{I}_N & g_v\textbf{I}_N
\end{bmatrix}, \;\;\;
\hat{\textbf{L}}=
\begin{bmatrix}
\textbf{L} & 0 \\
0 & \textbf{L}
\end{bmatrix}.
\end{equation*}
%and finally our extended Laplacian is
%\begin{equation*}
%\hat{\textbf{L}}=
%\begin{bmatrix}
%\textbf{L} & 0 \\
%0 & \textbf{L}
%\end{bmatrix}.
%\end{equation*}
Note here that the notation $\textbf{J}$ will be reserved to identify the Jacobian of the $2 \times 2$ reactions matrix:
\begin{equation*}
\textbf{J}=
\begin{bmatrix}
f_u  & f_v \\
g_u & g_v
\end{bmatrix}.
\end{equation*}
}

We then look for solutions to Eq.~\eqref{eq:turingPatternsLinearisedNetwork} of the form 
\begin{equation}
    \begin{aligned}
    \delta \textbf{u} = \sum_{\alpha = 1}^N b_\alpha e^{\sigma(\Lambda_\alpha) t}\boldsymbol{\Phi}^\alpha,\\
    \delta \textbf{v} = \sum_{\alpha = 1}^N c_\alpha e^{\sigma(\Lambda_\alpha) t}\boldsymbol{\Phi}^\alpha,
    \end{aligned}
\end{equation}
where $\Lambda_\alpha$, $\boldsymbol{\Phi}^\alpha$ are respectively the eigenvalues and eigenvectors of the Laplacian $\textbf{L}$ matrix, $\sigma(\Lambda_\alpha)$ are the eigenvalues of the extended Jacobian $(\hat{\textbf{J}}+ \textbf{D}\hat{\textbf{L}})$, and $\alpha$ is the index term.

\begin{figure*}[t!]
\centering
\includegraphics[width=1.02\textwidth]{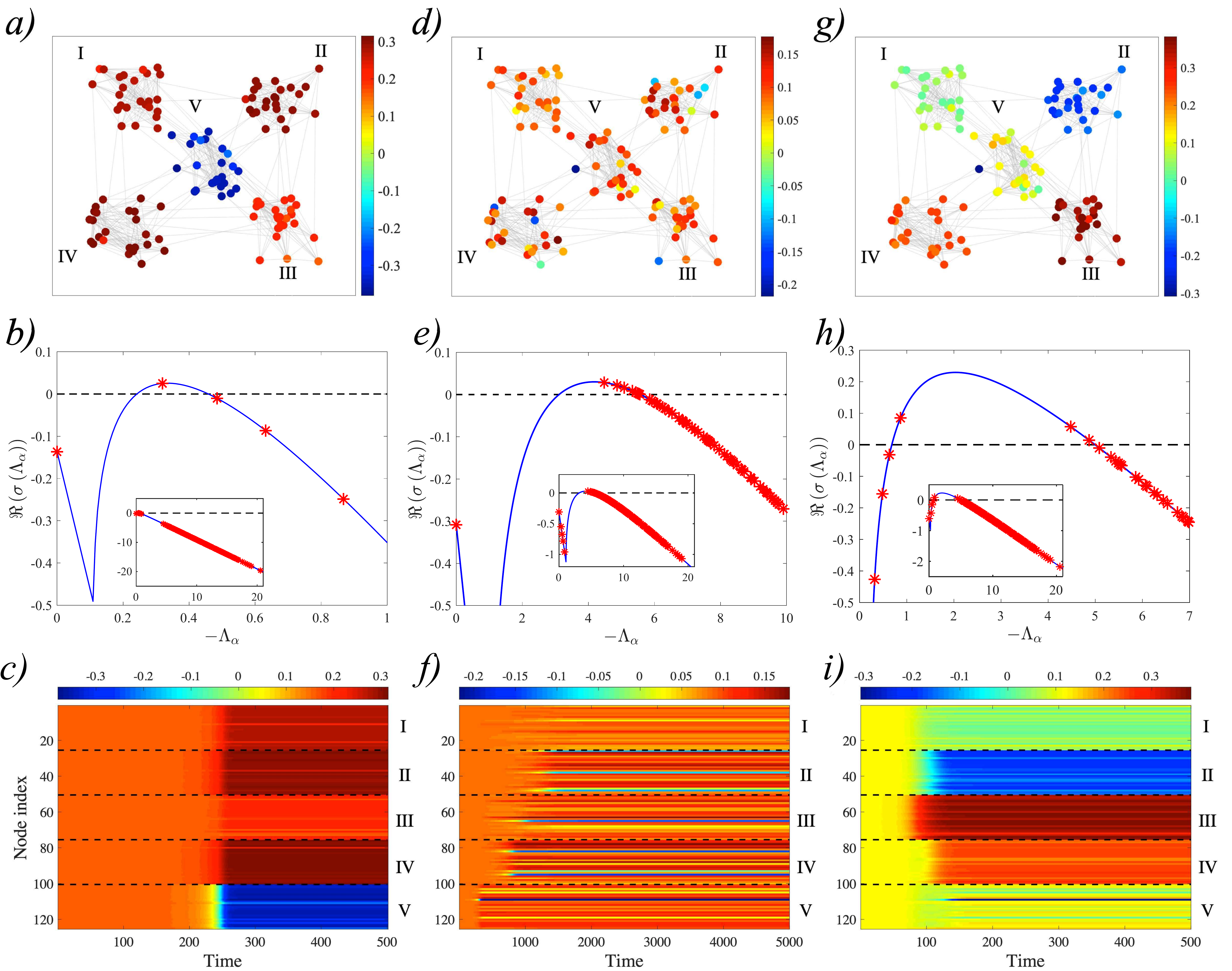}
\caption{\textbf{Patterns classification on modular networks.} $\textbf{a)}$ Modular patterns are formed when the concentration $u_i$ is homogeneous across all nodes in the same module. $\textbf{b)}$ In the corresponding dispersion relation we fix the parameters in order to have a single positive modular eigenvalue. The parameters are $a = 0.7$, $b = 0.05$, $c = 1.75$, $\rho = 5.5$, $D_u = 1$. $\textbf{c)}$ Temporal evolution of the modular pattern. $\textbf{d)}$A heterogeneous pattern emerges when the nodes inside the modules have different concentrations. $\textbf{e)}$ To show this, in the dispersion relation multiple non-modular eigenvalues are positive. The parameters are $a = 0.4$, $b = 0.05$, $c = 4.1$,  $\rho = 14$, $D_u = 0.1$. $\textbf{f)}$ Temporal evolution of the heterogeneous pattern, one can see that the instability first developed in the central module. $\textbf{g)}$ In-between pattern is a mixed state of the previous patterns. $\textbf{h)}$ In this case the instability comes from the contribution of both modular and non-modular eigenvalues. Here the parameters are $a = 0.6$, $b = 0.05$, $c = 3.625$, $\rho = 20$, $D_u = 0.16$. $\textbf{j)}$ Temporal evolution of the mixed pattern. In all the cases we used a network with $N=125$ nodes, $660$ edges, and a diameter of $d=5$. (Colour online)}
\label{fig:class_patt}
\end{figure*}

%Furthermore, we find that there are two sets of eigenvalues, those close to the origin which we denote as ``modular'' eigenvalues and those far from the origin, which we denote as ``non-modular'' eigenvalues. 
%The eigenvectors associated with the \textcolor{red}{modular} \textcolor{blue}{first set of} eigenvalues are organised on a per module basis; the \textcolor{blue}{remaining} eigenvector\textcolor{blue}{s are} \textcolor{red}{of the non-modular eigenvalues is} heterogeneous in a single module and close to zero everywhere else (see Fig.~\ref{fig:eigen_patt}).
As will be seen in the following, the description of the linear solution through the eigenvectors of the Laplacian matrix will be essential in our analysis for the prediction of the modularity of final nonlinear patterns. In fact, depending on which eigenvalues are positive, we can control the final shape of the pattern, as in Fig.~\ref{fig:class_patt}.

%Subbing
%\textcolor{magenta}{It may be worth introducing a reference here, since this is not a method that we came up with. e.g. ``Following the approach described in [citation], we substitute the expansion of perturbation into [equation reference], which decomposes...''}
Following the standard approach described by \cite{Murray2001,Othmer1971,Nakao2010}, we substitute the expansion of the perturbations into Eq.~\eqref{eq:turingPatternsLinearisedNetwork}. This decomposes the extended Jacobian to a $2\times 2$ matrix (for each index $\alpha$) for which the eigenvalue problem needs to be solved,
\begin{equation}
\textbf{J}_\alpha =
\begin{bmatrix}
f_u + D_u\Lambda_\alpha & f_v \\
g_u & g_v + D_v\Lambda_\alpha
\end{bmatrix},
\end{equation}
where subscripts on the activation function $f(u,v)$ and the inhibition function $g(u,v)$ represent partial derivatives evaluated at $(u^*,v^*)$ . 
To study the stability of the linear system we look for positive real parts of the eigenvalues of $\textbf{J}_\alpha$.
Turing instability occurs when the real part of the larger of the two eigenvalues $\sigma(\Lambda_\alpha) = \left(\text{tr}\textbf{J}_\alpha + \sqrt{(\text{tr}\textbf{J}_\alpha)^2 - 4 \text{det}\textbf{J}_\alpha}\right)\big/2$ is positive. 
relation between the eigenvalues of the extended Jacobian and the eigenvalues of the Laplacian, $\sigma(\Lambda_\alpha)$, is known in the  literature as the dispersion relation \cite{Murray2001}, for the continuous version see appendix \ref{subsec:appendixCOntinuousFormulation}.
For an activator-inhibitor system the necessary conditions for stability are $\text{tr}\textbf{J}_\alpha <0$ and $\text{det}\textbf{J}_\alpha > 0$. 
The first condition is always true, since $\text{tr}\textbf{J}_\alpha = \text{tr}\textbf{J} + (D_u+D_v)\Lambda_\alpha$, and this is negative since the stability of the fixed point in the absence of diffusion implies that $\text{tr}\textbf{J} <0$, while the non-positivity of the Laplacian spectrum implies $\Lambda_\alpha < 0$. 
We therefore turn our attention to the second condition for stability, which concerns $\text{det}\textbf{J}_\alpha=\text{det}\textbf{J} + \left(f_uD_v + g_vD_u\right) \Lambda_\alpha + D_uD_v\Lambda^{2}_\alpha$.
In order for a Turing instability to occur, we require $\text{det}\textbf{J}_\alpha < 0$.
Noting that the stability of the fixed point in the absence of diffusion implies that $\text{det}(\textbf{J}) > 0$ and noting that $\Lambda_\alpha < 0$, it is straightforward to conclude that the only way for $\text{det}\textbf{J}_\alpha $ to be negative is for $(f_u D_v + g_v D_u)$ to be positive.
%Following the same conditions as above for the trace, and noting that that the stability of  Turing patterns implies $\text{det}(\textbf{J}) > 0$, it is straightforward to conclude that the only way for $\text{det}\textbf{J}_\alpha $ to be negative is for $(f_u D_v + g_v D_v)$ to be positive. 
Without loss of generality we define $u$ to be the activator and $v$ to be the inhibitor.
Recalling the previous definition of an activator-inhibitor system, $u$ (respectively,  $v$) increases (respectively, decreases), the production of both species $u$ and $v$. As a result of this, the signs of the respective partial derivatives are $f_u > 0$, and $g_v < 0$.
Therefore, we require $\rho=D_v/D_u > 1$ for instability \cite{Turing1952,Othmer1971,Murray2001}, implying that the inhibitor should diffuse faster than the activator in order for Turing patterns to arise.
In many practical cases, this difference needs to be very large in order to achieve  $\det(\textbf{J}_\alpha)<0$.
%\textcolor{magenta}{Something is needed here in order to make the next section make more sense along the lines of ``Moreover, Turing instabilities will be easier to observe if $\rho$ is larger.'' I'm not 100\% comfortable with this because of the ``easier to observe'' phrase. Maybe instead we could say that there will be a critical value of $\rho$ above which Turing instabilities will arise?}

%\onecolumngrid        
        
\begin{figure*}[t]
\centering
\includegraphics[width=.85\textwidth]{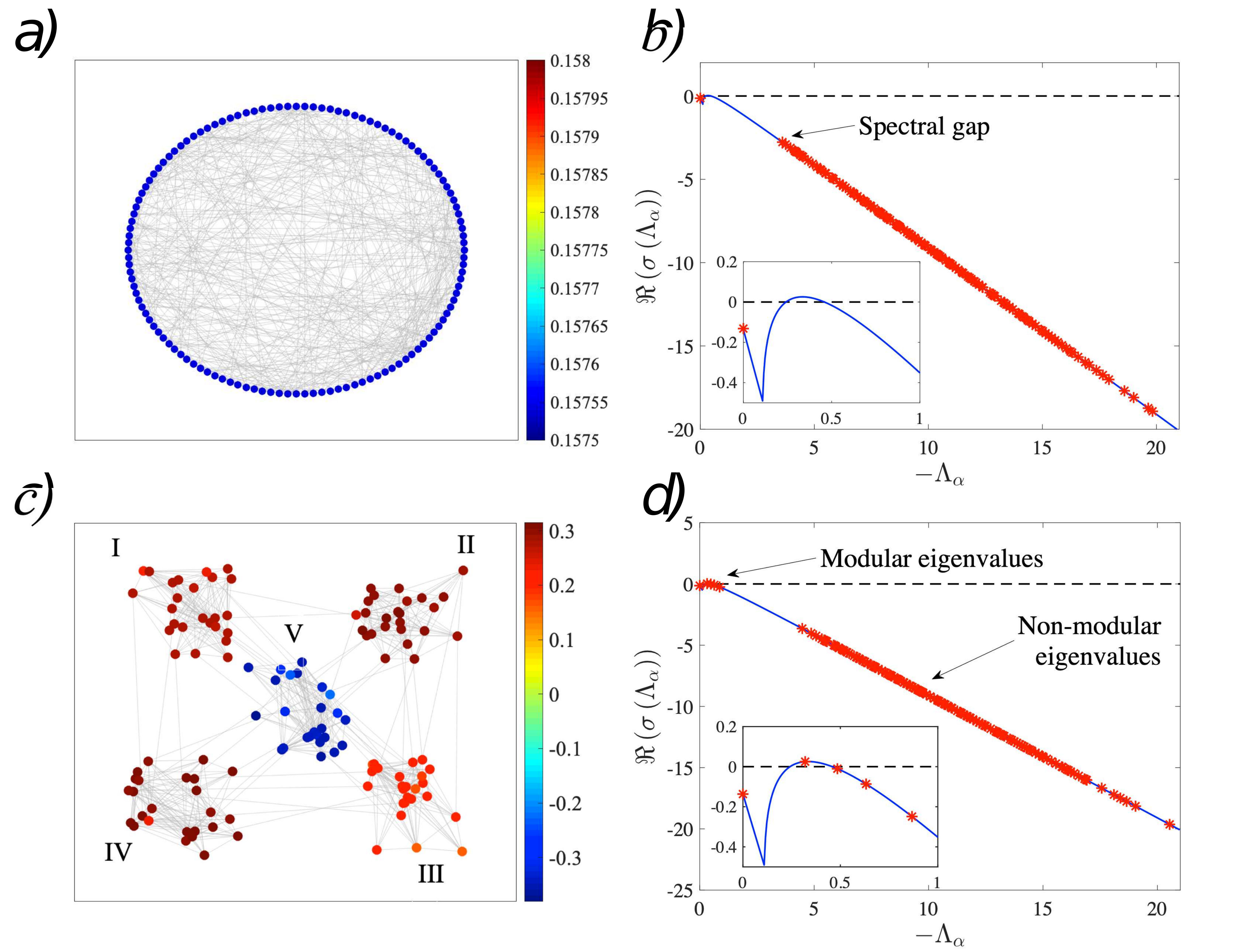}
\caption{\textbf{Modular vs. non-modular topology in Turing pattern formation.} 
%I have several thoughts on looking at these. I think the concentration colour-scales for a) and c) should be identical to make them comparable, and it may be worthwhile pointing out that $u^* \approx 0.157543$ (is there a reason for the discrepancy from this in (a) or just numerical error?). Is the diameter of the NW network specified or a consequence of the particular network generated? If the latter (as I would expect), it seems odd to mention it here. How long is long time? Do we define the ``dispersion relation of the continuous domain'' in the main text? Maybe be explicit about the fact that $\sigma(\Lambda)$ is defined for all $\Lambda$, but is only directly meaningful when $\Lambda$ is in the spectrum of the Laplacian? What do we mean by the modules being in the ER family? Surely that's inherent in the fact that the network was generated by a stochastic block model and we should just give the associated parameters? We talk about ``an important spectral gap'', but this is very different from the use of ``spectral gap'' in the main text. Lastly, it occurs to me that things would look a lot neater if we defined our Laplacian to be the negative of what we are currently using, just like spectral analysis of differential equations will typically use $-\nabla^2$ rather than $\nabla^2$ because $-\nabla^2$ is a nonnegative-definite operator, which has convenient consequences.
$\textbf{a)}$ A Newman--Watts (NW) network with $N=125$ nodes, and $660$ edges, where patterns are absent. The colour of the nodes represents the concentration of the activator, $u_i(t)$, at long time. $\textbf{b)}$ The dispersion relation of the NW network (red stars) overlain on the dispersion relation of the continuous case (blue curve), {i.e. if the system was on a continuous domain and not on a network, where we have subsituted the eigenvalues of the Laplacian with a wave number, $k^2$.} Notice the absence of the unstable eigenvalues (inset) and the gap between the zero eigenvalue and the second smallest $\Lambda_2$, known as the spectral gap. $\textbf{c)}$ A modular network of the same size (same number of nodes and edges) as in $a)$ where indeed Turing patterns are present. The five modules are of the Erd\H{o}s-R\'enyi (ER) family. The colour of the nodes again represents the concentration of the activator, $u_i(t)$, at long time. Note that the concentration of activator is homogeneous within modules, this is due to the modular nature of the network. $\textbf{d)}$ The dispersion relation of the modular network (red stars) overlain on the dispersion relation of the continuous domain (blue curve). Notice here the presence of unstable eigenvalues (inset) and that the eigenvalues are separated in two sets by an {important gap, between the first and second set of eigenvalues.} The first four non-zero eigenvalues are denoted as the modular eigenvalues and the remaining non-zero ones as the non-modular eigenvalues. The parameters of the FitzHugh--Nagumo model are in both cases $D_u = 1$, $\rho=5.5$, $a=0.7$, $b=0.05,$ $c=1.7$. {Finally, note the different colormaps used between panels $a)$ and $c)$ to highlight the lack of patterns in the former.}  (Colour online)}
\label{fig:Mod_vs_NW}
\end{figure*}
%\twocolumngrid

\subsection{The case for $D_v \gtrsim D_u$}
\label{subsec:DvDu}

%\textcolor{magenta}{I mostly like this section, although I will want to carefully reread the technical details once other fixes have been made. To avoid some ambiguity, I've introduced $\Lambda_\text{min}$ in place of $\Lambda_\alpha$ at some points. Another thing we might want to think about is the fact that $f_u(1+\epsilon)$ looks like $f_u$ is evaluated at $(1+\epsilon)$ not multiplied by it. Additionally, we should probably be explicit at the end of the section about how this all relates to the spectral gap. Be explicit that since $\Lambda_\text{min} \to 0$, it follows that a small spectral gap is needed to yield an eigenvalue of the Laplacian that corresponds to a positive eigenvalue of the Jacobian.}
From experimental observations~\cite{Horsthemke1987,Pearson,exp_Turing1,exp_Turing2} it is rarely true  that the inhibitor diffuses much faster than the activator, but instead the chemicals diffuse with similar rates. %, making Turing theory limited to only a small portion of the parameter space. 
In the case where $D_v \gtrsim D_u$, it can be shown that the dispersion relation is positive only for values of the spectrum of the Laplacian very near to the origin. 
To prove this we analyse the behaviour of $\text{det}(\textbf{J}_\alpha)$ when considered as a function of $\Lambda_\alpha$;
more precisely, we focus on the value of $\Lambda_\alpha$ corresponding to a minimum of $\text{det}(\textbf{J}_\alpha)$. {It is known in literature \cite{Murray2001} that for the continuous case, it will always exist a non-positive value of $\Lambda_\alpha$ such that the $\text{det}(\textbf{J}_\alpha)<0$ or, in other words, that Turing instability can occur. In order to proceed with our analysis, in the following, we will consider that $\Lambda_\alpha$ takes continuous values and will see that the spectrum of a (strongly) modular network} fits in the domain of the continuous dispersion relation for which the instability occurs for the particular case, $D_v \gtrsim D_u$. We start by differentiating with respect to $\Lambda_\alpha$ and after some algebraic manipulation, we find that the minimum of $\text{det}\textbf{J}_\alpha$ is found at $\Lambda_\alpha = \Lambda_\text{min}$ where
\begin{equation}
\Lambda_\text{min} = -\frac{f_u\rho + g_v}{2D_v}.
\label{eq:rho=1}
\end{equation}
From relation~\eqref{eq:rho=1} we note that if $D_v$ is kept fixed while $\rho \rightarrow 1$ then $\Lambda_\text{min}\rightarrow 0$. 
To show this we set $\rho = 1 + \epsilon$.
%\textcolor{blue}{Assuming $\Lambda_\text{min}$ is non-positive, $(1 + \epsilon)f_u + g_v >0$.
%Rearranging, we can write $(1 + \epsilon)f_u + g_v = \text{tr}\textbf{J}+\epsilon f_u$ and, noting that $\text{tr}\textbf{J}$ is necessarily negative, we conclude that the positive quantity $\text{tr}\textbf{J}+\epsilon f_u$ shrinks with smaller values of epsilon.
%Thus the numerator of \eqref{eq:rho=1} also shrinks with $\epsilon$.
%%In order to enforce $\tetbf{detJ}_\alpha < 0$, $\Lambda_\alpha$ must be between 0 and -2\frac{\epsilon f_u + \textbf{trJ}}{D_v} \cite{Murray2001}
%Therefore, as $\epsilon$ get smaller, the value of $\Lambda_\alpha$ for which $\text{det}\textbf{J}_\alpha$ is at its minimum tends towards the origin.
%Hence, the possible values of $\Lambda_\alpha$ that may permit Turing instabilities tend towards zero as the ratio $\rho$ of diffusivities tends to $1$.}
Under the conditions of the Turing instability, $\Lambda_\text{min}$ is non-positive, so $(1 + \epsilon)f_u + g_v >0$.
Rearranging, we can write $(1 + \epsilon)f_u + g_v = \text{tr}\textbf{J}+\epsilon f_u$ and, noting that $\text{tr}\textbf{J}$ is necessarily negative, we conclude that the positive quantity $\text{tr}\textbf{J}+\epsilon f_u$ can be at most of order $\epsilon$, since $\epsilon f_u > |\text{tr}\textbf{J}|$. 
%, we see that 
%$ 0 > \text{tr}\textbf{J} > -\epsilon f_u $.
This shows that $\Lambda_\text{min}$ is of order $\epsilon$.
Therefore, as $\epsilon$ decreases, the value of $\Lambda_\alpha$ for which $\text{det}\textbf{J}_\alpha$ is at its minimum tends towards the origin.
Hence, the possible values of $\Lambda_\alpha$ {that may permit Turing instabilities tend towards zero as the ratio $\rho$ of diffusivities tends to $1$.
%Therefore the maximum point of the dispersion relation is found for values of $\Lambda_\alpha$ closer to zero.
In practice, this implies that the range of values of $\Lambda_\alpha$ for which instabilities can occur decreases in size and is restricted to small values of $\Lambda_\alpha$}.
Therefore, a small spectral gap is needed to allow patterns to form.
{This is significant for the analysis of modular networks that follows since, as shown in the following section, modular networks are characterised by a small spectral gap $\lvert \Lambda_2 - \Lambda_1 \rvert$.
Hence the Laplacian of a modular network will have eigenvalues close to the origin.
Because of this, we are able to find modular networks where Turing instabilities, and thus pattern formation, may occur where otherwise (i.e., in non-modular networked systems) they would not. 
This modular pattern formation may even occur for values of $\rho$ that are close to those observed in real systems.}

%\onecolumngrid            
\begin{figure*}[t!]
    \centering
    \includegraphics[width=1.02\textwidth]{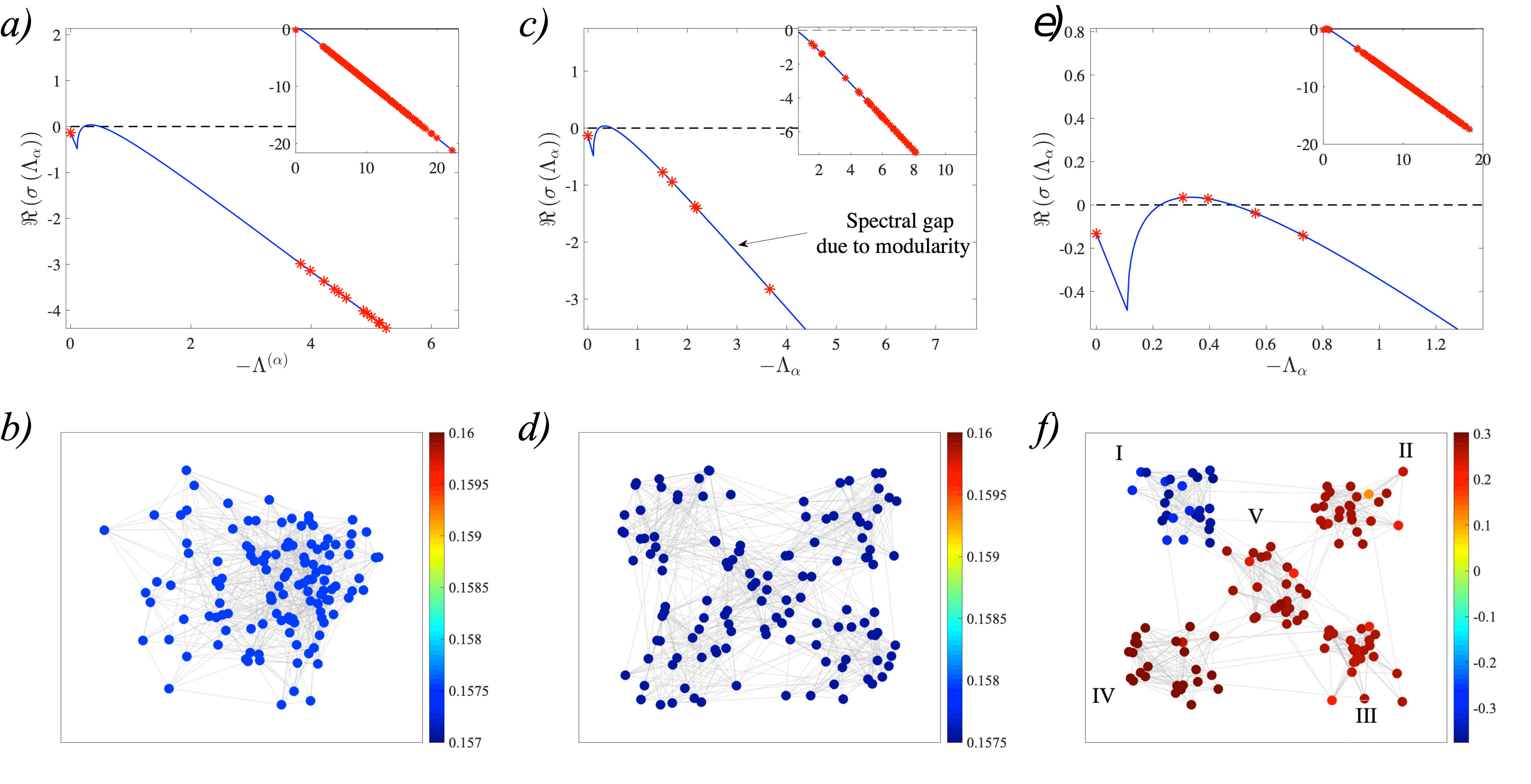}
    \caption{\textbf{Emergence of patterns by changing the modularity.} $\textbf{a)}$ The dispersion relation for an ER network (shown in panel $\textbf{b)}$), made of $125$ nodes and $660$ edges, and modularity measure $Q = 0.02562$. $\textbf{c)}$ The dispersion relation for a weakly modular network shown in panel $\textbf{d)}$ consisting of $5$ modules and $540$ intra-edges within modules, and $120$ inter-edges between modules, and modulalarity $Q = 0.6150$. Notice that there is an emerging gap now between the first $4$ non-zero eigenvalues and the rest of them. $\textbf{e)}$ 
    %And finally 
    The dispersion relation for a strongly modular network $\textbf{f)}$ with $630$ intra-edges, $30$ inter-edges, and $Q = 0.7545$. Notice that the spectral gap between the zero eigenvalue and the smallest non-zero eigenvalue is much smaller and a pattern has formed on the network. For all simulations $D_u = 1$, $\rho = 5.5$, $a = 0.7$, $b = 0.05$, and $c = 1.7$. Also we used the algorithms described in Refs. \cite{NewmanFiedler,Reichardt2006} with resolution parameter $\gamma = 1$ the modularity $Q$ in each case.  (Colour online)}
    \label{fig:changingModularity}
\end{figure*}

%\twocolumngrid  

\section{Turing patterns on modular networks}
\label{sec:whymodular}

It has been argued that that the existence of particular topological features in many types of networks, including brain networks, are of crucial importance in several important processes from neuronal communication \cite{Sporns_Zwi} to structural robustness \cite{albert_barabsi}.
Such functional properties are based on the short average path length that characterises this family of networks. 
We emphasised in the preceding subsection that the spectral gap is an important ingredient for the Turing instability. 
In this section, we further illustrate  this fact by taking into account a special family of networks, the modular ones, that are known for for their lack of spectral gap. As a comparison we contrast the process of pattern formation in a non-modular network such as a Newman--Watts (NW) network (a particular case of a small-world network) with the pattern formation on a modular network generated using the Stochastic Block Model (SBM).

As described in \cite{SpornsBook,Sporns2004,Meunier2010}, modular structure has been identified in many brain networks.
Since the FitzHugh--Nagumo model~\cite{FitzHugh1961, Nagumo1962} is both useful for modelling neuronal dynamics ~\cite{Murray2001}, and since it can exhibit spatial pattern formation \cite{Murray2001,Asllani2014Directed}, we will use this model throughout this paper.
In dimensionless form, FitzHugh--Nagumo dynamics correspond to using the functions $f(u,v)= u - u^3 - v$ and $g(u,v) = c(u - a + bv)$ to describe the net production of activator and inhibitor in Eq.~(\ref{eq:turingPatternNetworks}) where $a$, $b$, and $c$ are constants.
The parameters of the model are always chosen such that we have a stable fixed point.

%\textcolor{magenta}{There is a big jump here! `Comparison' could sound like looking at the network topologies rather than comparing the results of FitzHugh--Nagumo simulations. We need to say that we did numerical solutions of the FitzHugh--Nagumo model on different networks, as well as computing the spectral properties of the Laplacian and Jacobian. We also need to be more explicit about the construction of the networks and the performance of the simulations (possibly with reference to an appendix). Was the NW network built from a 2-nearest-neighbour ring and then 160 edges added, or from a 1-nearest-neighbour ring and then 410 edges added? Do the networks have precisely 660 edges---and if so, how was this achieved---or an expected value of 660 edges? What are the probabilities used in the SBM presented? What initial conditions were used in the FitzHugh--Nagumo simulations? In general, it is a good policy to give the reader enough information that they could replicate our results. While not as important, it would also be good to be explicit as to whether these were the only simulations tried or whether these are indicative simulations from a larger family (with different sizes, different average degrees, etc.) and whether we saw similar results.}
In Fig.~\ref{fig:Mod_vs_NW} we { compare the pattern on a} single-module NW network (of $125$ nodes and $660$ edges) and a modular network with $5$ communities, each with $25$ nodes and a {local} Erd\H{o}s--R\'enyi (ER) topology. 
As can be observed from the dispersion relation in Fig.~\ref{fig:Mod_vs_NW} b), the distribution of the eigenvalues of the Laplacian matrix for the NW network, which shows a large spectral gap. 
This makes the Turing instability impossible for the given choice of parameters (including $\rho = 5.5$), since the instability (i.e., values of $\Lambda$ corresponding to positive values of the continuous curve) is concentrated %in the part of the spectrum 
near the origin. 
We could potentially create an instability by significantly increasing $\rho$, or optimise the rewiring to minimise the diameter.
%textcolor{magenta}{In the introduction, we are explicit with the idea that a pattern is a macroscopic structure that is the distinctive mark of self-organisation. Since we are talking about the ``null pattern'' here, maybe a different word should be used?}
As $t \to \infty$, the FitzHugh--Nagumo models considered in this paper will tend to an equilibrium. 
One way to depict these equilibria is to plot the concentration of the activator species at long times. 
For the non-modular network described above, this is shown in Fig.~\ref{fig:Mod_vs_NW} a) and we see that the activator concentration is homogeneous across all nodes as expected.
%\textcolor{magenta}{Avoid the word groups unless it's being used with its technical meaning!! Also, the sentence needs finessing for grammar, clarity, and linking; notice that the previous sentence was about the concentrations of activator at equilibrium, and now we are returning to talking about the spectrum. Perhaps start a new paragraph, which could begin: ``In contrast to the large spectral gap observed in the NW network, we find that there are several eigenvalues of the Laplacian of the modular network that are located near the origin. Specifically, if $M$ is the number of modules we find that there the Laplacian has $M-1$ nonzero eigenvalues located near the origin. This is consistent with the spectral theory of modular networks [citation, maybe to Peixoto: DOI:10.1103/PhysRevLett.111.098701, which could also be useful earlier].''}

In contrast to this, for a strongly modular topology the spectrum is divided into two distinct sets of eigenvalues.
The first set is those nonzero eigenvalues near the origin (of which there are $M-1$ where $M$ is the number of the modules) and the second set is composed of all the remaining eigenvalues that are far from the origin \cite{Peixoto}.
{We note that both the NW network and the modular network have the same number of nodes and edges, so the difference between the networks' spectra cannot be attributed to a difference in the number of nodes or in the average degree of these nodes.}
As already anticipated, we will refer to the first  set of {nonzero} eigenvalues of the Laplacian matrix as the modular eigenvalues (for example in Fig.~\ref{fig:Mod_vs_NW} d) the first four non-zero eigenvalues). 
{In Fig.~\ref{fig:Mod_vs_NW} d), we observe that the modular eigenvalues  are sufficiently close to 0 and in the interval of possible values of the spectrum where the instability can develop; in Fig.~\ref{fig:Mod_vs_NW} c) we see that this leads to a pattern in the activator concentrations at equilibrium.}

%\textcolor{magenta}{New idea, new paragraph? A lot of this section seems to be strung together without paragraphing. And I'm starting to have doubts about how this paper is structured. Perhaps it would be useful to do all of the spectral theory together, perhaps between presenting the theory of Turing instabilities on networks (and thus demonstrating the relevance of spectral theory) and before presenting the results of FitzHugh--Nagumo simulations. Also, why are we concentrating on NW networks? There is a large class of networks that will have a small diameter (ER graphs, for instance!) and that seems to be the key topological property in the analysis that follows. Maybe we can make broader statements rather than concentrating on NW networks, even if we prefer to present our results on NW networks. TBH, it is still unclear to me why we are presenting results on NW networks rather than ER graphs in this paper...}
To understand the reason why the spectrum of a modular network can be divided into two subsets we should first  explain the reason behind the spectral gap in small-world networks.
As mentioned earlier, the denomination ``small-world'' {refers to a certain class of networks, one feature of which is the small average distance between nodes.}
In \cite{Bojan1991}, Bojan shows that the absolute value of the second largest Laplacian eigenvalue $\lvert\Lambda_2\rvert$ is bounded below by $\big\lvert \dfrac{4}{Nd} \big\rvert$, where $N$ is the number of nodes in the network and $d$ is the diameter.
This means that for a fixed value of the size $N$ of the network, the lower bound of the spectral gap (equivalently, $|\Lambda_2|$) is larger when the diameter $d$ is smaller; impeding this way a non-modular network like the NW under consideration, having a smaller spectral gap than a modular network.~\footnote{blueWe want to emphasise that regular networks (e.g., rings) have a large diameter, too, having this way a small spectral gap. However, our focus here is on random networks which, apart from the modular ones, are characterised by a small diameter.}

%\textcolor{magenta}{This is the paragraph that will lead into the plots demonstrating what happens as the parameters of the stochastic block model are changed. It ultimately needs to be refashioned to make this its aim or (alternatively, if the paper is restructured to introduce a section on the spectral properties associated with different network topologies) we need to give a paragraph in the ``spectral properties'' section about the spectra of modular networks and a paragraph in the ``numerical results'' section about the patterns observed on modular networks with different parameters in the stochastic block model. The most important thing that needs to be added is a quantification of the different stochastic block models (which will go with the fact that we need an explicit description of how the SBMs are constructed in an earlier section) and how the parameters of an SBM relate to its spectral properties.}
To further investigate how the spectral gap changes for different network topologies, we look at three different networks in Fig. \ref{fig:changingModularity}.
We create these networks in a simple way. {First we divide our $125$ nodes into five modules of nodes, and define the total number of intra-edges (connections within modules) and inter-edges (connections between modules).
{Then we allocate each module an equal number of intra-edges and inter-edges and randomly connect nodes within and between the modules, while avoiding double entries in both cases.}
%Then we allocate each module one fifth of the total inter-edges, and use these to randomly connect nodes between modules, once again avoiding double entries.
If we define the number of intra-edges to be much larger than the number of inter-edges, then this process will yield a network with as strong modular structure.}
We describe three examples of these networks with increasing ``modularity'', where modularity is defined by the $Q$ function described in \cite{NewmanFiedler,Reichardt2006}.
We first look at an ER graph, as shown in Fig. \ref{fig:changingModularity} $b)$.  %\ref{subsec:appendixHierarchy}
Notice that there is a large spectral gap in the corresponding dispersion relation, as shown in Fig. \ref{fig:changingModularity} $a)$.
{By simply modifying the ratio of inter-edges to intra-edges, we can then generate a new network which begins to close the spectral gap, as in Fig. \ref{fig:changingModularity} $c)$ and $d)$.}
Finally in Fig. \ref{fig:changingModularity} $f$) we have reduced the number of inter-edges such that patterns form, and the spectral gap is greatly decreased, as in Fig. \ref{fig:changingModularity} $e)$.
Notice again that there are two set of eigenvalues, the first four non-zero eigenvalues (which we refer to as modular eigenvalues) and the remaining non-modular eigenvalues.
This leads us to ask why a highly modular network closes the spectral gap so well.
{Note that in the Appendix, we consider the hierarchical case where each module is arranged in a small-world fashion.}

To understand the small spectral gap of modular networks, we first imagine a scenario in which the modules are disconnected from each other. 
Individually, these modules are denser and smaller than the Newman--Watts network, therefore each of them is expected to have a relatively large spectral gap. 
From the algebraic connectivity theorem \cite{Golub} we know that the number $M$ of the connected components (the modules in this case) corresponds with the number of zero eigenvalues of the Laplacian operator. 
%\textcolor{magenta}{We mention a ``spectrum perturbative analysis''. Should there be a reference to a paper, or a more explicit description of this here, or is the spectrum perturbative analysis an informal observation based on the idea that the spectrum of the Laplacian should not change dramatically as links are added. If the last of these, is there a neat justification for the fact that the spectrum doesn't change too much? It's plausible and reasonable, but I don't know how it's justified.}
However, once the modules are connected with a small number of links then $M-1$ of these eigenvalues will move away from zero. From a spectrum perturbative analysis, we find that these become very small nonzero eigenvalues, with only one zero eigenvalue still remaining to signify that the whole modular network is connected. \footnote{ Notice also that due to the algebraic connectivity, a network with a Laplacian spectral gap will always be modular.} 
This explains the small size of the spectral gap in modular networks and consequently the emergence or not of Turing patterns respectively in small-world and modular networks{ \cite{Donetti,Andreotti}.}   

\begin{figure*}[t]
\centering
\includegraphics[width=.85\textwidth]{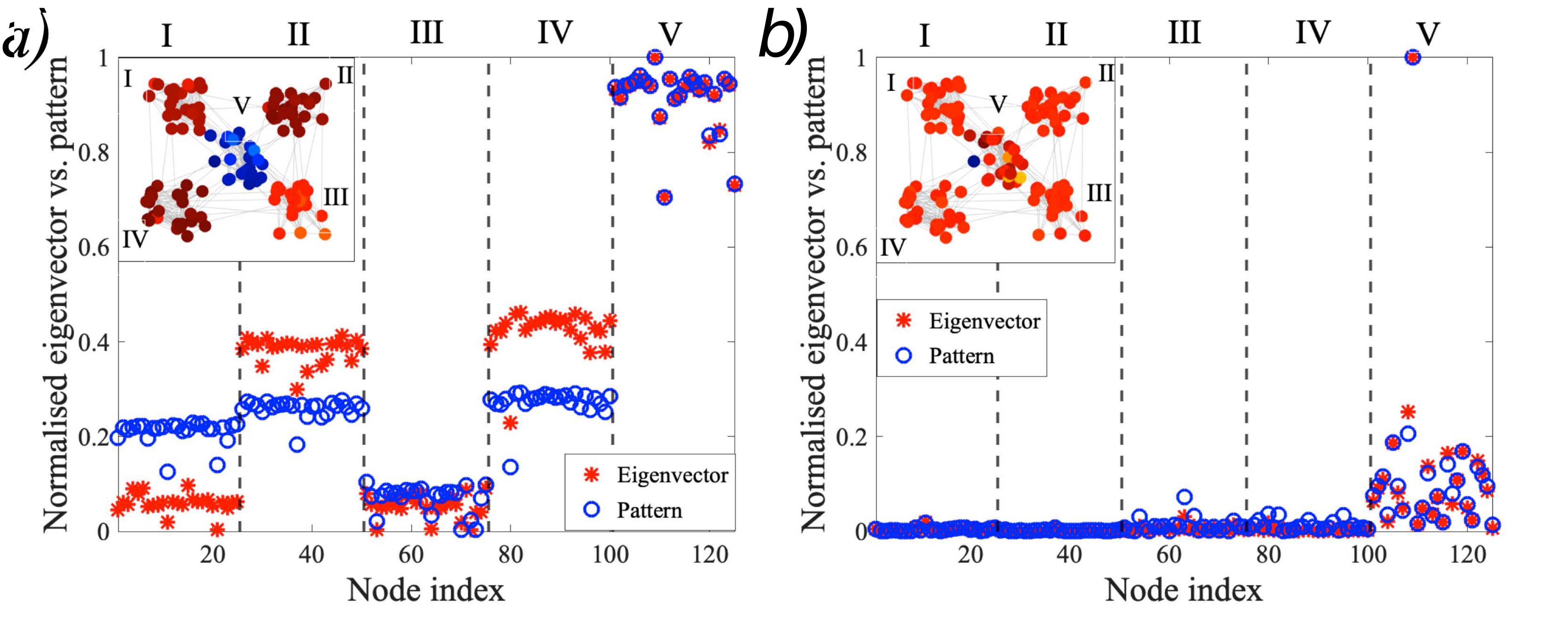}
\caption{\textbf{Origin of modular patterns.} In this figure, we plot the normalised patterns versus the normalised eigenvectors corresponding to a single positive modular and non-modular eigenvalue, respectively. $\textbf{a)}$ A modular pattern (inset) is formed when one (or more) of the set of the modular eigenvalues is unstable and dominant over the non-modular ones. 
The parameters are the same as in Fig.~\ref{fig:class_patt} $a)$. $\textbf{b)}$ However, when the non-modular eigenvalues dominate over the rest of the spectrum then heterogeneous pattern are created (inset). Furthermore in this setting it is also possible to identify the origin of instability from which the pattern first emerges, in this case the central module.%\textcolor{magenta}{How is the normalisation performed? The ``normal'' (haha) way of normalising an eigenvector would be by its magnitude, but here the normalisation appears to be by its largest component (infinity-norm). How are the equilibrium patterns normalised? Presumably this is associated with the discrepancy from the homogeneous equilibrium value? If instead we're currently using an affine transformation to normalise the smallest value to 0 and the largest value to 1, this would easily explain why Fig a) is off for the interior modules. It would be useful to fix this. Note also that it looks like there is a sign change hidden somewhere in Figure b): Node 100-and-something has a highly negative value at equilibrium, but a highly positive value in the normalised version?}
The normalisation for the patterns is simply $\dfrac{|u_i - u^*|}{\max_i|u_i-u^*|}$ for each entry $i$, and the same normalisation is used for the eigenvector. Also the nodes are organised here in blocks of $25$ individuals for each module. Finally the parameters for $b)$ are $D_u = 0.1$, $\rho = 13.38$, $a = 0.4$, $b = 0.05$, $c = 4$ (Colour online).}
\label{fig:eigen_patt}
\end{figure*}

We notice from Figure~\ref{fig:Mod_vs_NW} $c)$ that although the pattern is highly heterogeneous at a global level, the patterns on nodes within each single module are quite homogeneous, having almost the same concentration of the species for each node in the module. 
Such macroscopic spatially extended patterns where densely connected entities (e.g., of biological nature) show the same amount of activity have been observed in different biological contexts \cite{NewmanPNAS,metabolic,protein} and in particular in dynamics of the brain \cite{Smith2018, Oldguys}. While \cite{Smith2018} is mainly an experimental paper, and first highlights the observation of spatial patterns on brain networks, we have laid down a rigorous mathematical foundation that explores the importance of modularity to the formation of Turing patterns. Additionally, to the best of our knowledge, we here propose the first self-organising mechanism that explains the uniformity at the module level of Turing patterns in biological networks.

We can obtain insight into the patterns of $u$ and $v$ observed at equilibrium by constructing and analysing the eigenvectors associated with the Turing instabilities. From an initial condition close to the unstable homogeneous equilibrium, the rate of change in the concentrations $u$ and $v$ will initially be dominated by the eigenvector associated with the largest positive eigenvalue of the Jacobian. This initial growth will ultimately be stabilised by nonlinear terms, and we expect that the state equilibrium pattern of concentrations will be reminiscent of the eigenvectors associated with the instability \cite{Turing1952,Murray2001}.
%\textcolor{magenta}{What does ``To start with'' mean here? What are we starting? And how do we select a single modular eigenvalue to be unstable? Since this is not something that we can specify, what do we modify in order to construct a network where a single modular eigenvalue is unstable? I suspect a different phrasing is needed here. Also, I don't think we have yet discussed the fact that the eigenvectors associated with the modular eigenvalues are ``modular''. This is another observation that weighs in favour of including a section on spectral theory before a section presenting the results of Turing instability experiments.}
To begin our analysis of the resultant patterns, we select parameters which lead to a single modular eigenvalue being positive, and observe the final ``homogeneous by module'' pattern as in Fig.~\ref{fig:class_patt} $a)$, $b)$, and $c)$. 
The situation changes when the instability is exclusively induced from the non-modular eigenvalues. 
In this case the concentration is no longer uniform for each module as shown in Fig.~\ref{fig:class_patt} $d)$, $e)$, and $f)$.
A hybrid state is obtained instead when both sets of eigenvalues contribute to the Turing instability as in Fig.~\ref{fig:class_patt} $g)$, $h)$, and $i)$. 
These hybrid states can lead to patterns that are similar to either the modular patterns or the heterogeneous patterns. This is because the Turing instability in this case involves a competition between the eigenvectors associated with the unstable modular eigenvalues and the eigenvectors associated with the non-modular eigenvalues. The dominant instability (and therefore the eigenvector that we expect to be most similar to the equilibrium pattern) will be the eigenvector associated with the largest eigenvalue of the Jacobian. In Fig.\ \ref{fig:class_patt} \emph{h)} for example, we observe that the largest eigenvalue of the Jacobian is associated with one of the modular eigenvalues of the Laplacian, and this is associated with a pattern in Fig.\ \ref{fig:class_patt} \emph{i)} that could be described as almost being modular.
In the Supplementary Material (SM) we discuss several criteria to establish which eigenvalue is dominating over the others. 

%\onecolumngrid            

\begin{figure*}[t]
\centering
\includegraphics[width=1.02\textwidth]{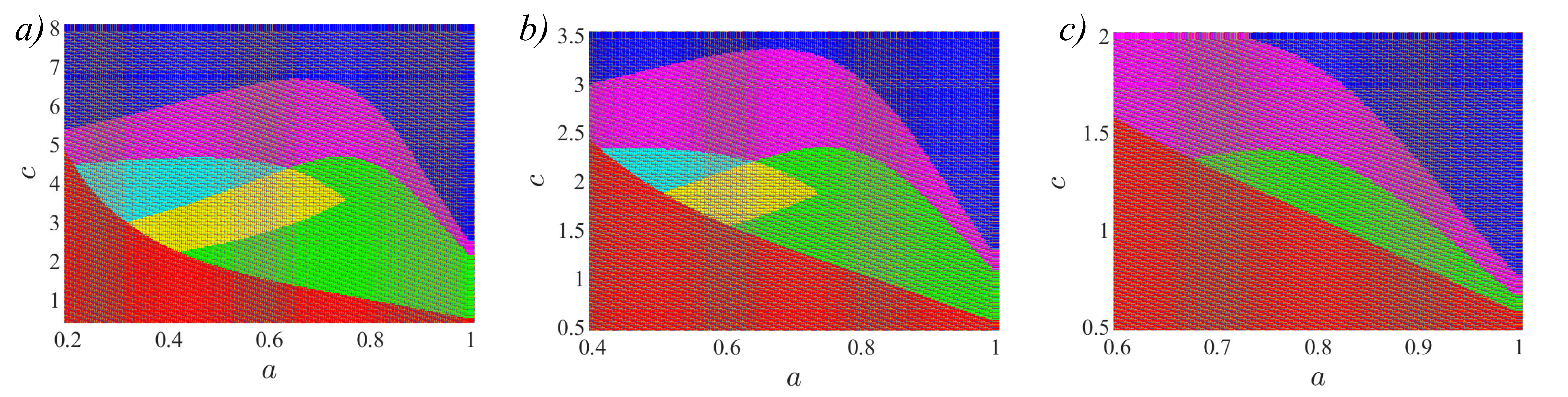}
\caption{\textbf{Parameter space for decreasing diffusivities ratio.} 
We classify different types of pattern on modular graphs in the parameter space of the FitzHugh--Nagumo model $(a,c)$, a fixed value of $b= 0.05$, $D_u = 0.15$, and $a)$ $\rho = 20$, $b)$ $\rho = 10$, $c), \rho = 6$. The portion of the parameter space indicated in red represents the region where no Turing patterns are allowed, as the system (in the absence of diffusion) is not in a steady state. All other regions correspond to  parameters sets for which Turing instability is allowed. The blue part is when the system is Turing stable, that is the system is at a steady state but no Turing patterns form. The rest of the region is when patterns may occur: in the magenta region patterns only form in the continuous domain case, in green we have ``modular'' patterns, Fig. \ref{fig:class_patt} $a)$, yellow ``mixed'' state patterns, Fig. \ref{fig:class_patt} $g)$, and cyan heterogeneous patterns, Fig. \ref{fig:class_patt} $d)$. Notice that as the ratio of diffusivites approaches 1, $\rho\rightarrow 1$, the only patterns which form are the modular patterns, showing that in a real scenario modularity is the only way to induce pattern formation in networks with modular structures, e.g., brain networks.  (Colour online)}
\label{fig:param_space}
\end{figure*}
%\twocolumngrid     

In order to understand why the final shape of the pattern can be modular we focus on the study of the eigenvectors as plotted in Fig.~\ref{fig:eigen_patt}.
%\textcolor{magenta}{Why would the stability analysis tell us that the pattern from the instability is retained at the stable equilibrium? Also this paragraph is full of ``shapes of eigenvectors'' that make me feel uncomfortable.}
From the stability analysis we know that initially the pattern is shaped according to the unstable eigenvectors and this form is largely retained in the final nonlinear regime. 
Nevertheless, what surprises is the particular form of the eigenvectors associated with the modular eigenvalues as in Fig.~\ref{fig:eigen_patt} $a)$; in particular, the fact that the components of the modular eigenvectors are segregated accordingly to the respective modules \footnote{ Notice here that it may be, as in the case of Fig. \ref{fig:eigen_patt}, that different modules might share by chance the same level of components. However, this should not be understood as these entries belonging to the same module.}.  
To shed light on this peculiarity we will resort again to  spectral graph theory. 

As anticipated earlier, the smallest non zero eigenvalue of the Laplacian $\Lambda_2$ defines the spectral gap known also in the literature as the Fiedler eigenvalue and defines the algebraic connectivity \cite{Fiedler, Chung}.
%\textcolor{magenta}{A citation or more detailed description is needed for this property of the Fiedler vector. Can we be more explicit here on what is meant by ``similar values''?}
Its corresponding eigenvector is known as the Fiedler eigenvector and has the property that the entries of the nodes corresponding to the same modules take very similar values. 
Because of this property, the Fiedler eigenvector has been extensively used as the basis of several community detection methods \cite{NewmanFiedler,Donetti,Andreotti}.
However it should be noted that the Fiedler partitioning can underestimate the total number of modules as we show in Fig.~\ref{fig:eigen_patt}.
The other modular eigenvectors also behave in a similar manner to the Fiedler eigenvector; their entries are segregated by module \cite{Donetti,Andreotti}. 
%In this sense considering the complexity of modular eigenvectors 
{Since the modular eigenvectors are often the fastest growing modes in the Turing instability, this means that the modular} {shape} of the global pattern is a consequence of the modularity of the structure of the network itself. 

On the other hand, when the instability is caused strictly by the non-modular eigenvalues, another behaviour occurs during the pattern forming phenomenon.
%\textcolor{magenta}{The following sentence (and indeed the rest of the paragraph) does not make sense to me. I think what we are trying to say is again about the spectral properties of modular graphs (and could therefore go in a section on the spectral properties of modular graphs). I suspect that the point is the following (and feel free to use this if it is correct): If a graph $G$ consists of disconnected subgraphs, then the nonzero components of any eigenvector associated with a nonzero eigenvalue of the Laplacian of $G$ will correspond to nodes in the same module [citation/clarification needed, because I'm pretty sure I can construct a counterexample to this]. Introducing a small number of additional edges to $G$ will not make dramatic changes to the eigenvalues or eigenvectors [citation needed]. Thus, the largest (in magnitude) components of a nonmodular eigenvector of the Laplacian of a modular graph will correspond to nodes in a single module. If this nonmodular eigenvector is associated with a Turing instability, it will lead to a pattern where the deviations away from the homogeneous equilibrium occur mainly in a single module. (???)}
{This is best considered by again considering a modular network to be a perturbation of a network with initially $M$ disconnected components. In such a case, each nonzero eigenvalue of the Laplacian will correspond to an eigenvector whose components are all zero outside a single component. A modular network will be a small perturbation to this, and so the non-modular eigenvectors will also be close to zero except within a single component. If only one non-modular eigenvalue corresponds to a Turing instability, then only one module of the network will show pattern formation, as illustrated in Fig.~\ref{fig:eigen_patt} b). Thus, we can predict the module on which pattern formation will occur by looking at the components of the eigenvector whose eigenvalue corresponds to the fastest growing mode of the Turing instability.}
%\textcolor{blue}{
%The non-modular eigenvalues belong to different components, therefore their eigenvectors only contribute to the pattern on their component.
%Thus, if the only positive eigenvalue is a non-modular eigenvalue, then only the corresponding component will form a pattern on it Fig.~\ref{fig:eigen_patt} $b)$. 
%Even if the components are loosely connected, the perturbation of the eigenvector is still comparatively small.
%Therefore, we can predict, by looking at the eigenvector of the largest positive eigenvalue, the component from which the pattern originates.}
%Since the non-modular eigenvalues belong to different components when the modules are supposed disconnected even when they are attached weakly (with few links) with each other their corresponding eigenvectors will be slightly perturbed. 
%In this way, if a non-modular eigenvalue is unstable it will activate only the corresponding  module because the eigenvector will be different from zero (almost) only in the entries of the corresponding module as in Fig.~\ref{fig:eigen_patt} $b)$, so identifying the region of the network (brain) where the patterns originates. 

So far we have considered the contribution in the formation of patterns of both modular and non-modular eigenvalues, however when we deal with Turing patterns in real scenarios the ratio $\rho=D_v/D_u$ is quite close to one \cite{Horsthemke1987,Pearson,exp_Turing1,exp_Turing2}.
To evaluate the conditions under which different patterns form in real conditions we now explore the parameter space of the FitzHugh--Nagumo dynamics in more detail.
Note that in order for a Turing pattern to form, we must begin from a stable fixed point.

%\textcolor{magenta}{Is it clear why it is most relevant to vary $a$ and $c$ while keeping $b$ and $D_u$ fixed? If we have performed the numerical experiments for other changes but conclude that the changing $b$ and $D_u$ are uninteresting, it would be good to say this. Have we nondimensionalised to determine how many truly independent parameters there are in this system? I suspect that we have (and that this is why the equation for $u$ has no free parameters), but it would be good to check this. The following sentence also needs editing.}
In Fig.~\ref{fig:param_space} it can be observed that although different types of patterns can be found in the space of the parameters $a$ and $c$, as the ratio of diffusivities gets closer to 1 the region where patterns can occur shrinks and, more importantly, the only possible Turing patterns are modular ones (indicated in green colour).
One could find patterns in the other regions by tuning the diffusion parameters, except in the red region due to the absence of a stable fixed point.

%\textcolor{magenta}{In what sense to brain networks optimise their spatial interaction matrix and what justification do we have for the claim that they do this in order to allow pattern formation?}
{The result that brain networks have optimised their spatial interaction matrix in order to allow pattern formation has been already claimed by experimental observers \cite{Smith2018,Oldguys}; to the best of our knowledge we present the first mechanism that explains the role of modularity in achieving this pattern formation.}   
        
\begin{figure*}[t]
\centering
\includegraphics[width=.9\textwidth]{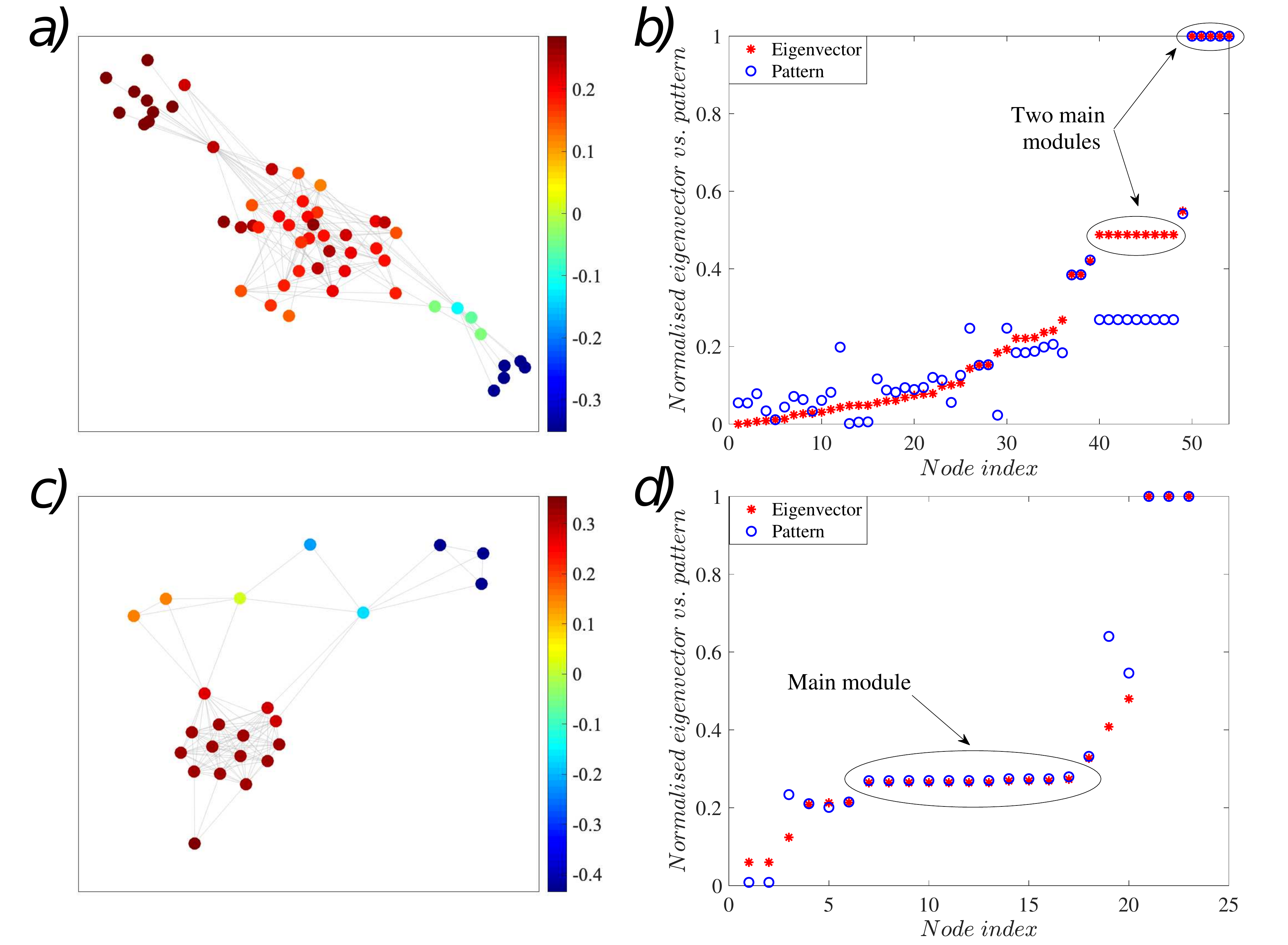}
\caption{\textbf{Modular patterns on real-world networks} $\textbf{a)}$ The modular pattern of the neuronal network of $54$ nodes of the nematode \textit{P. Pacificus} \cite{ppacificus} with parameters $D_u = 0.7$, $\rho = 4.5$, $a = 0.75$, $b = 0.04$, $c = 1.5$. $\textbf{b)}$ The comparison between the normalised unstable eigenvector and the final pattern shows the presence of two distinct modules. $\textbf{c)}$ The modular pattern of the network of $23$ individuals (nodes) of a zebra herd \cite{zebra} with parameters $D_u = 0.7$,  $\rho = 4.5$, $a = 0.75$, $b = 0.04$, $c = 1.5$. $\textbf{d)}$ The comparison between the normalised unstable eigenvector and the final pattern shows the presence of a main module. Note that, in order to concentrate on the effects of modularity only, in both cases the networks have been simplified to be undirected, and unweighted. We have also extracted the giant component in each case. (Colour online)
%$e)$ The modular pattern (of a disease spreading, see Refs. \cite{Sun,Sun_rev}) on a road transportation network of Chicago $823$ nodes \cite{konectLink, konectArticle1, konectArticle2, konectProceedings} with parameters $D_u = 0.7$, $\sigma = 4.5$, $a = 0.75$, $b = 0.04$, $c = 1.5$. $d)$ The comparison between the normalised unstable eigenvector and the final pattern shows the presence of multiple modules. In all the cases the shape of the patterns fits well the theoretical prediction.
}
\label{fig:C_elegans}
\end{figure*}

\section{Self-organisation in real modular networks}
\label{sec:realNetworks}    
Heretofore we have discussed the role of modularity in the formation of patterns only for synthetic networks.
In this part we will illustrate our findings in real examples of biological or ecological networks. 
The neuronal networks of several primitive animals such as nematodes have been well characterised. Indeed, it was the study of nematode neuronal networks that first inspired the development of small-world network models \cite{Watts1998}.
In Fig.~\ref{fig:C_elegans} $a)$ we show the final modular pattern of the nematode \textit{P. pacificus} \cite{ppacificus}.
This follows from the theoretical prediction of the unstable Fiedler eigenvector, shown in Fig.~\ref{fig:C_elegans}$b)$
Here we have used the Fiedler eigenvector to identify the communities of neurons \cite{NewmanFiedler}.
In this particular case two modules are clearly distinguishable and the level of activity of the nodes inside the modules are quite homogeneous. Other examples of Turing patterns in neuronal networks are presented in the Supplementary Material. Although the modularity of brain networks has been well-studied \cite{SpornsBook,Meunier2010,Sporns2004,Sporns_Zwi} other types of natural networks manifest this property also. For instance, this is the case for ecological networks where the individuals are connected to each other through trophic relations \cite{Murray2001,zebra}. 
Such modular contact networks have also been shown to be crucial for the pattern of disease spreading \cite{Sun,Sun_rev}.
In Fig. \ref{fig:C_elegans} $c)$ and $d)$ we present respectively the {equilibrium} pattern {of the FitzHugh--Nagumo equations} and its comparison to the unstable eigenvector of {the contact network of} a zebra herd \cite{zebra} where a community of $11$ individuals out of a total of $23$ is clearly visible.
However, the formation of patterns of spreading are not limited only to contact networks, which in general can be small in size. 
Modularity is a common property in other types of networks which, although they are not directly related to biological systems, are still essential for biological phenomena occurring on them.
This is for instance, the case for networks of human mobility, such the roads networks in the city of Chicago presented in the SM \cite{konectLink, konectArticle1, konectArticle2, konectProceedings}, which are decisive for the spreading of an epidemics in the entire urban area \citep{Sun,Sun_rev}.
%In the panels $e)$ and $f)$ is shown the organisation of this network in $6$ main function units, which have different but homogeneous per module amount of concentrations of the susceptible and infected individuals. 
These examples all show agreement with the mathematical analysis we have shown so far.

\section{Discussion and conclusions}

In this paper we have analytically and numerically explored pattern formation on modular networks. 
We have shown that modularity, a ubiquitous topological feature of many biological networks, is crucial for the self-organisation of the global dynamics on a network.
To study this behaviour we have considered here the Turing instability as a paradigmatic mechanism for pattern formation in biology, ecology or neuroscience. 
{The possibility of pattern formation via the Turing mechanism} on non-modular networks is limited to unrealistically extreme ratios of the diffusion constants of the activator and inhibitor species making the small spectral gap of the Laplacian matrix a fundamental requirement for the Turing instability. 
%\textcolor{magenta}{I would take a few more sentences to state this next part clearly since it is the main point of the paper.}
This feature is a structural advantage of modular networks which follows from spectral perturbation theory.
A strongly modular network can be considered as a set of connected components weakly attached  with {a small number of intermodule links. From spectral perturbation theory this yields a number -- equal to one fewer than the number of modules -- of non zero eigenvalues very near to the origin.} 
This characterisation at the linear stability level influences the shape of the spatially extended patterns. 
Due to the segregation of the entries of the eigenvectors corresponding to the set of modular eigenvalues, we are able to explain why Turing patterns are homogeneous per module on these networks.
%Due to the segregation that the entries of the eigenvectors (e.g. Fiedler eigenvector) corresponding to the emergent modular eigenvalues, we are able to explain why Turing patterns are modular, namely homogeneous per module, in the natural scenario. 

This result opens to an important aspect regarding the functional resolution of the brain modes which was hypothesised \cite{Sporns2016, Hutt,Meunier2010} in several experimental observations \cite{Smith2018, Oldguys}.
To the best of our knowledge, the model we present here constitutes the first self-organising mechanism where the modules are presented as functional blocks of biological networks. 
In this sense, we argue that the module is the smallest spatial unit to be taken into account from the functional point of view {i.e. if we ``zoom'' out far enough from a modular network, the individual modules behave like individual supernodes.}
For the particular example of the brain the modules might be the super-nodes of the functional connectomes \cite{Sporns2004,Sporns2016,Smith2018}. 
%In this sense, 
Indeed, the (self-)segregation of the network structure in modules \cite{NewmanPNAS} influences also the shape of the dynamical pattern on it. 
Based on the fact (see \cite{Meunier2010,Hutt,Sporns2004} and Fig.\ref{fig:C_elegans}) that in real scenarios Turing patterns should be exclusively modular, we believe that the results we have shown here can be potentially used to formulate a community detection protocol \cite{NewmanPNAS,NewmanFiedler,NewmanBook} in the case where patterns of self-organised activity are known to exist.

In the case when we relax Turing conditions to allow the instability for the non-modular part of the spectrum, then we can {use the eigenvector corresponding to the largest eigenvalue to indicate} the module in which the Turing pattern is first seeded before finally spreading to the rest of the network. 
This behaviour can potentially make the pattern formation process a powerful diagnostic tool for studying and eventually controlling the emergence of abnormal dynamics which characterise many neurological diseases \cite{AsllaniPLOS} or the spread of an epidemic in a group of individuals \cite{Sun,Sun_rev}. 
{We test our theoretical results on several real connection data sets of neuronal, ecological and infrastructure networks verifying the correctness of our findings, that modularity is crucial for the development of patterns, and that when the instability is derived from the first set of modular eigenvalues, that the resultant self-organisation follows the modular structure of the network.}  

The results we have presented here can extend also to more complicated scenarios. This is, for example, the case when the hierarchy of a network is considered as a complement to its modularity. In the Appendix we show that in a hierarchical modular network the modular eigenvalues are even more relevant for the Turing pattern forming process.
Further extensions of our approach are also possible; for example
%Following also the same track that we have explored here
%, it is of particular interest exploring also more complex topologies of modular networks where for instance directed networks are considered. 
to consider the effect of directed edges in a modular network. In this case we expect a richer dynamics where travelling Turing waves should emerge in a directed modular networks \cite{Asllani2014Directed}.

\section*{Acknowledgements}
B.~A.~Siebert acknowledges funding from the Irish Research Council under grant GOIPG/2018/3026. The  work of J.~P.~Gleeson and M.~Asllani is partly funded by Science Foundation Ireland (grant numbers 16/IA/4470, 16/RC/3918, 12/RC/2289 P2, 18/CRT/6049) and co-funded under the European Regional Development Fund.

\section*{Appendix}
\subsection{The FitzHugh--Nagumo model}

{We have used the Fitzhugh--Nagumo model throughout this paper \cite{FitzHugh1961, Nagumo1962},} which is one of the first and best-known mathematical models used to describe the spiking dynamics of neurons. 
In terms of mathematical equations the behaviour of a single neuron is described by
\begin{eqnarray}
\frac{d u}{d t} &=& u - u^3 - v\nonumber\\
\frac{d v}{d t} &=& c(u - av - b)
\end{eqnarray}
where $u$ is the membrane potential and $v$ the recovery variable. The model itself was first introduced by FitzHugh \cite{FitzHugh1961} to explain the generation of spikes in excitable systems, i.e.,  neurons. A spike is a short-lasting elevation of the membrane voltage $u$ diminished over time by a slower and linear recovery variable $v$ once the system is periodically excited by an external current. The following year Nagumo \textit{et al.} \cite{Nagumo1962} developed the electric circuit which mimics such behaviour. 
However, although the model itself is mainly used to describe the oscillatory behavior of neurons, it also admits a stable fixed point, which is a necessary requirement {for Turing instabilities}.
Once this model is equipped with a diffusion term, it turns out in a perfect candidate for pattern formation \cite{Murray2001}. In recent years, with the rapid development of network science, the FitzHugh--Nagumo model has been extended to diffusively coupled networks \cite{Perc2005, Asllani2014Directed}.

\subsection{Continuous Formulation}
\label{subsec:appendixCOntinuousFormulation}
The original continuous framework for pattern formation, in one dimension, is
\begin{equation}
\begin{aligned}
\frac{\partial u}{\partial t} &=& f(u,v) + D_u \frac{\partial^2u}{\partial x^2},\\
\frac{\partial v}{\partial t} &=& g(u,v) + D_v \frac{\partial^2v}{\partial x^2},
\end{aligned}
\end{equation}
where notation is as in Sec.~\ref{sec1}.
The derivation of Turing patterns follows the same process as we describe in the main text, except in a continuous form.
Of note, the extended Jacobian is now
    \begin{equation}
        \textbf{J}_k = \textbf{J} - \textbf{D}k^2 =
        \begin{bmatrix}
            J_{11} - k^2 D_u & J_{12}\\
            J_{21} & J_{22} - k^2 D_v
        \end{bmatrix}
    \end{equation}
where $k$ is the wave number.
Then, the continuous dispersion relation, $\sigma(k)$, is plotted against the wave number, $k^2$ instead of the eigenvalues of the Laplacian.
\subsection{The role of hierarchy of the brain networks in the pattern formation}
\label{subsec:appendixHierarchy}        
\begin{figure*}[t]
\centering
\includegraphics[width=.9\textwidth]{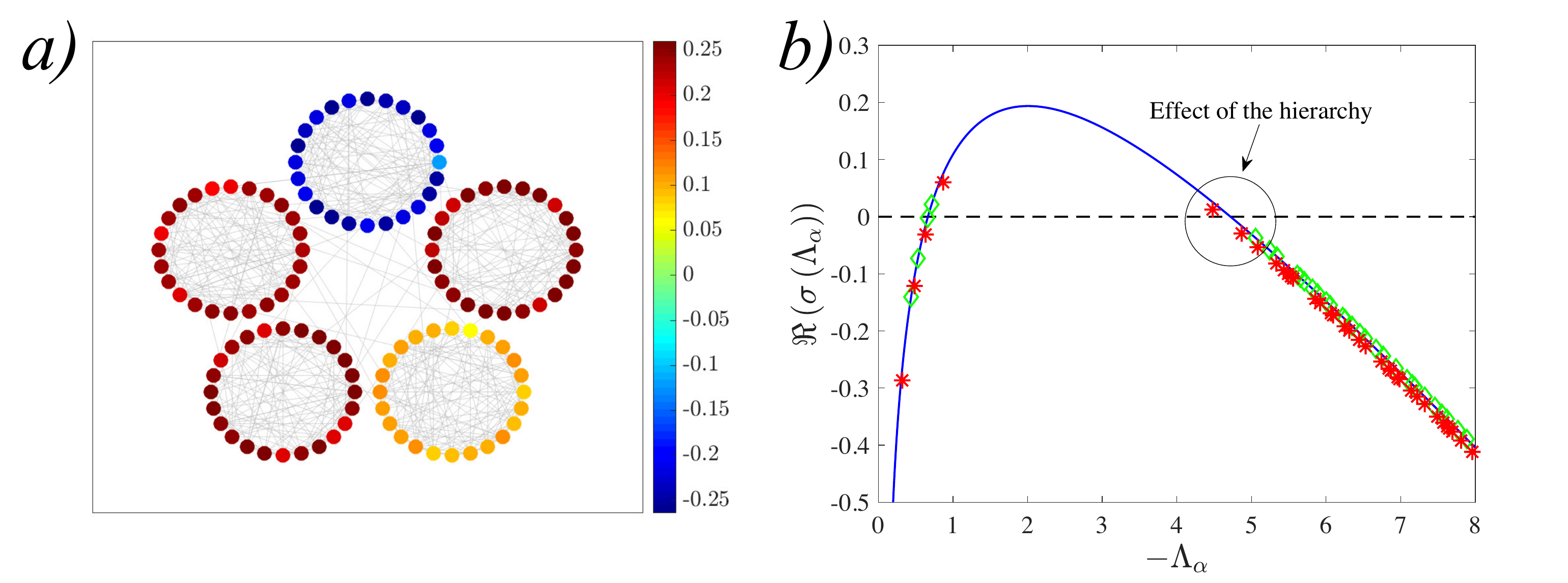}
\caption{\textbf{Patterns on hierarchical modular networks.} $\textbf{a)}$ The modular pattern on five connected Newman--Watts networks, each with $N=25$ nodes and $125$ edges. Each module is connected to its clockwise and counterclockwise neighbours with two random edges. There are a further $25$ random edges added to induce shortcuts, as in the Newman--Watts style. $\textbf{b)}$ The dispersion relation of the hierarchical network (green diamonds) and modular network with random (ER) modules (red stars) with the same number with $D_u = 0.15$, $\rho =17$, $a=0.7$, $ b=0.05$, $c=3.625$. The larger gap between the first five eigenvalues and the rest arises due to the hierarchy of the network.  (Colour online)}
\label{fig:hierar_mod}
\end{figure*}

We have discussed the role that modularity has on pattern formation, isolating it from other topological features, which is in fact an integral aspect of many networks, including brain networks.
So then a question that arises naturally is, how does the brain cope with maintaining both features and their functional properties at the same time? 
We now are able to answer this question by recalling an important empirical results that characterises most real networks, their hierarchical structure \cite{hierarchical,NewmanBook,Meunier2010}. 
In fact, most of the connectomes studied are organised in a modular structure, however each module is further organised in a small-world fashion. This is another amazing observation how nature tends to self-organize in order to better optimise the benefit from the both structural features, the modularity and the small-worldness.
In a hierarchical modular network the entire network is organised in modules which are attached to each other so as to have a small diameter and at the same time the nodes in the modules are connected in such way to form sub-modules again minimising their diameter and this process goes on this way up to smallest building unity, the single nodes.
A hierarchical structure stresses once more the necessity of modularity for the self-organising phenomena in the networks. 
In Fig.~\ref{fig:hierar_mod} we show that the difference of the smallest non-modular eigenvalue from the origin is larger when the modules have a small-world topology compared to when they are organised at random (e.g. ER network) for the same number of nodes, edges and modules. 
The reason for this can be found once more by taking a perturbative approach.
The spectral gap of an individual module (disconnected from the rest of the network) is larger when its diameter is smaller, as it is in the Newman--Watts network used in Fig.~\ref{fig:hierar_mod}.
%This reason for that can be searched once more in the perturbative approach, the spectral gap of each module thought as disconnected from each other is larger when the diameter is smaller, as it is in a Newman--Watts network used in Fig.~\ref{fig:hierar_mod}.

Thus, in the presence of hierarchy, the cyan and the yellow regions in Fig.~\ref{fig:param_space} would be even smaller making the modularity region shown in green larger compared to the previous two.
We notice, however, that the instability invariance is still valid for values of the diffusivites ratio $\rho$ near to $1$, that is when only the green region in the parameter space is available.   
In conclusion, a hierarchical arrangement where each module is arranged in a small world fashion, and these modules are again connected in a small world fashion, are even better candidates for forming modular patterns than the modular networks studied in the main text.

\newpage

\bibliography{mybib.bib}

\begin{thebibliography}{67}%
\makeatletter
\providecommand \@ifxundefined [1]{%
 \@ifx{#1\undefined}
}%
\providecommand \@ifnum [1]{%
 \ifnum #1\expandafter \@firstoftwo
 \else \expandafter \@secondoftwo
 \fi
}%
\providecommand \@ifx [1]{%
 \ifx #1\expandafter \@firstoftwo
 \else \expandafter \@secondoftwo
 \fi
}%
\providecommand \natexlab [1]{#1}%
\providecommand \enquote  [1]{``#1''}%
\providecommand \bibnamefont  [1]{#1}%
\providecommand \bibfnamefont [1]{#1}%
\providecommand \citenamefont [1]{#1}%
\providecommand \href@noop [0]{\@secondoftwo}%
\providecommand \href [0]{\begingroup \@sanitize@url \@href}%
\providecommand \@href[1]{\@@startlink{#1}\@@href}%
\providecommand \@@href[1]{\endgroup#1\@@endlink}%
\providecommand \@sanitize@url [0]{\catcode `\\12\catcode `\$12\catcode
  `\&12\catcode `\#12\catcode `\^12\catcode `\_12\catcode `\%12\relax}%
\providecommand \@@startlink[1]{}%
\providecommand \@@endlink[0]{}%
\providecommand \url  [0]{\begingroup\@sanitize@url \@url }%
\providecommand \@url [1]{\endgroup\@href {#1}{\urlprefix }}%
\providecommand \urlprefix  [0]{URL }%
\providecommand \Eprint [0]{\href }%
\providecommand \doibase [0]{http://dx.doi.org/}%
\providecommand \selectlanguage [0]{\@gobble}%
\providecommand \bibinfo  [0]{\@secondoftwo}%
\providecommand \bibfield  [0]{\@secondoftwo}%
\providecommand \translation [1]{[#1]}%
\providecommand \BibitemOpen [0]{}%
\providecommand \bibitemStop [0]{}%
\providecommand \bibitemNoStop [0]{.\EOS\space}%
\providecommand \EOS [0]{\spacefactor3000\relax}%
\providecommand \BibitemShut  [1]{\csname bibitem#1\endcsname}%
\let\auto@bib@innerbib\@empty
%</preamble>
\bibitem [{\citenamefont {Nicolis}\ and\ \citenamefont
  {Prigogine}(1977)}]{Nicolis1977}%
  \BibitemOpen
  \bibfield  {author} {\bibinfo {author} {\bibfnamefont {G.}~\bibnamefont
  {Nicolis}}\ and\ \bibinfo {author} {\bibfnamefont {I.}~\bibnamefont
  {Prigogine}},\ }\href@noop {} {\emph {\bibinfo {title} {{Self-Organization in
  Nonequilibrium Systems. From Dissipative Structures to Order Through
  Fluctuations.}}}}\ (\bibinfo  {publisher} {Wiley},\ \bibinfo {year}
  {1977})\BibitemShut {NoStop}%
\bibitem [{\citenamefont {Murray}(2001)}]{Murray2001}%
  \BibitemOpen
  \bibfield  {author} {\bibinfo {author} {\bibfnamefont {J.~D.}\ \bibnamefont
  {Murray}},\ }\href@noop {} {\emph {\bibinfo {title} {{Mathematical Biology II
  : Spatial Models and Biomedical Applications}}}}\ (\bibinfo  {publisher}
  {Springer-Verlag},\ \bibinfo {year} {2001})\BibitemShut {NoStop}%
\bibitem [{\citenamefont {Turing}(1990)}]{Turing1952}%
  \BibitemOpen
  \bibfield  {author} {\bibinfo {author} {\bibfnamefont {A.~M.}\ \bibnamefont
  {Turing}},\ }\href@noop {} {\bibfield  {journal} {\bibinfo  {journal}
  {Bulletin of Mathematical Biology}\ }\textbf {\bibinfo {volume} {52}},\
  \bibinfo {pages} {153} (\bibinfo {year} {1990})}\BibitemShut {NoStop}%
\bibitem [{\citenamefont {Gierer}\ and\ \citenamefont
  {Meinhardt}(1972)}]{Gierer1972}%
  \BibitemOpen
  \bibfield  {author} {\bibinfo {author} {\bibfnamefont {A.}~\bibnamefont
  {Gierer}}\ and\ \bibinfo {author} {\bibfnamefont {H.}~\bibnamefont
  {Meinhardt}},\ }\href@noop {} {\bibfield  {journal} {\bibinfo  {journal}
  {Biological Cybernetics}\ }\textbf {\bibinfo {volume} {12}},\ \bibinfo
  {pages} {30} (\bibinfo {year} {1972})}\BibitemShut {NoStop}%
\bibitem [{\citenamefont {Othmer}\ and\ \citenamefont
  {Scriven}(1971)}]{Othmer1971}%
  \BibitemOpen
  \bibfield  {author} {\bibinfo {author} {\bibfnamefont {H.~G.}\ \bibnamefont
  {Othmer}}\ and\ \bibinfo {author} {\bibfnamefont {L.~E.}\ \bibnamefont
  {Scriven}},\ }\href@noop {} {\bibfield  {journal} {\bibinfo  {journal}
  {Journal of Theoretical Biology}\ }\textbf {\bibinfo {volume} {32}},\
  \bibinfo {pages} {507} (\bibinfo {year} {1971})}\BibitemShut {NoStop}%
\bibitem [{\citenamefont {Schnabel}\ \emph {et~al.}(2006)\citenamefont
  {Schnabel}, \citenamefont {Bischoff}, \citenamefont {Hintze}, \citenamefont
  {Schulz}, \citenamefont {Hejnol}, \citenamefont {Meinhardt},\ and\
  \citenamefont {Hutter}}]{Schnabel2006}%
  \BibitemOpen
  \bibfield  {author} {\bibinfo {author} {\bibfnamefont {R.}~\bibnamefont
  {Schnabel}}, \bibinfo {author} {\bibfnamefont {M.}~\bibnamefont {Bischoff}},
  \bibinfo {author} {\bibfnamefont {A.}~\bibnamefont {Hintze}}, \bibinfo
  {author} {\bibfnamefont {A.-k.}\ \bibnamefont {Schulz}}, \bibinfo {author}
  {\bibfnamefont {A.}~\bibnamefont {Hejnol}}, \bibinfo {author} {\bibfnamefont
  {H.}~\bibnamefont {Meinhardt}}, \ and\ \bibinfo {author} {\bibfnamefont
  {H.}~\bibnamefont {Hutter}},\ }\href@noop {} {\bibfield  {journal} {\bibinfo
  {journal} {Developmental Biology}\ }\textbf {\bibinfo {volume} {294}},\
  \bibinfo {pages} {418} (\bibinfo {year} {2006})}\BibitemShut {NoStop}%
\bibitem [{\citenamefont {Holland}\ and\ \citenamefont
  {Hastings}(2008)}]{Holland2008}%
  \BibitemOpen
  \bibfield  {author} {\bibinfo {author} {\bibfnamefont {M.~D.}\ \bibnamefont
  {Holland}}\ and\ \bibinfo {author} {\bibfnamefont {A.}~\bibnamefont
  {Hastings}},\ }\href@noop {} {\bibfield  {journal} {\bibinfo  {journal}
  {Nature}\ }\textbf {\bibinfo {volume} {456}},\ \bibinfo {pages} {792}
  (\bibinfo {year} {2008})}\BibitemShut {NoStop}%
\bibitem [{\citenamefont {Horsthemke}\ \emph {et~al.}(2004)\citenamefont
  {Horsthemke}, \citenamefont {Lam},\ and\ \citenamefont
  {Moore}}]{Horsthemke2004}%
  \BibitemOpen
  \bibfield  {author} {\bibinfo {author} {\bibfnamefont {W.}~\bibnamefont
  {Horsthemke}}, \bibinfo {author} {\bibfnamefont {K.}~\bibnamefont {Lam}}, \
  and\ \bibinfo {author} {\bibfnamefont {P.~K.}\ \bibnamefont {Moore}},\
  }\href@noop {} {\bibfield  {journal} {\bibinfo  {journal} {Physics Letters
  A}\ }\textbf {\bibinfo {volume} {328}},\ \bibinfo {pages} {444} (\bibinfo
  {year} {2004})}\BibitemShut {NoStop}%
\bibitem [{\citenamefont {Nakao}\ and\ \citenamefont
  {Mikhailov}(2010)}]{Nakao2010}%
  \BibitemOpen
  \bibfield  {author} {\bibinfo {author} {\bibfnamefont {H.}~\bibnamefont
  {Nakao}}\ and\ \bibinfo {author} {\bibfnamefont {A.~S.}\ \bibnamefont
  {Mikhailov}},\ }\href@noop {} {\bibfield  {journal} {\bibinfo  {journal}
  {Nature Physics}\ }\textbf {\bibinfo {volume} {6}},\ \bibinfo {pages} {544}
  (\bibinfo {year} {2010})}\BibitemShut {NoStop}%
\bibitem [{\citenamefont {Asllani}\ \emph
  {et~al.}(2014{\natexlab{a}})\citenamefont {Asllani}, \citenamefont
  {Busiello}, \citenamefont {Carletti}, \citenamefont {Fanelli},\ and\
  \citenamefont {Planchon}}]{Asllani2014Multiplex}%
  \BibitemOpen
  \bibfield  {author} {\bibinfo {author} {\bibfnamefont {M.}~\bibnamefont
  {Asllani}}, \bibinfo {author} {\bibfnamefont {D.~M.}\ \bibnamefont
  {Busiello}}, \bibinfo {author} {\bibfnamefont {T.}~\bibnamefont {Carletti}},
  \bibinfo {author} {\bibfnamefont {D.}~\bibnamefont {Fanelli}}, \ and\
  \bibinfo {author} {\bibfnamefont {G.}~\bibnamefont {Planchon}},\ }\href@noop
  {} {\bibfield  {journal} {\bibinfo  {journal} {Physical Review E -
  Statistical, Nonlinear, and Soft Matter Physics}\ }\textbf {\bibinfo {volume}
  {90}},\ \bibinfo {pages} {1} (\bibinfo {year}
  {2014}{\natexlab{a}})}\BibitemShut {NoStop}%
\bibitem [{\citenamefont {Asllani}\ \emph
  {et~al.}(2014{\natexlab{b}})\citenamefont {Asllani}, \citenamefont
  {Challenger}, \citenamefont {Pavone}, \citenamefont {Sacconi},\ and\
  \citenamefont {Fanelli}}]{Asllani2014Directed}%
  \BibitemOpen
  \bibfield  {author} {\bibinfo {author} {\bibfnamefont {M.}~\bibnamefont
  {Asllani}}, \bibinfo {author} {\bibfnamefont {J.~D.}\ \bibnamefont
  {Challenger}}, \bibinfo {author} {\bibfnamefont {F.~S.}\ \bibnamefont
  {Pavone}}, \bibinfo {author} {\bibfnamefont {L.}~\bibnamefont {Sacconi}}, \
  and\ \bibinfo {author} {\bibfnamefont {D.}~\bibnamefont {Fanelli}},\
  }\href@noop {} {\bibfield  {journal} {\bibinfo  {journal} {Nature
  Communications}\ }\textbf {\bibinfo {volume} {5}},\ \bibinfo {pages} {1}
  (\bibinfo {year} {2014}{\natexlab{b}})}\BibitemShut {NoStop}%
\bibitem [{\citenamefont {Asllani}\ \emph {et~al.}(2015)\citenamefont
  {Asllani}, \citenamefont {Busiello}, \citenamefont {Carletti}, \citenamefont
  {Fanelli},\ and\ \citenamefont {Planchon}}]{Asllani2015}%
  \BibitemOpen
  \bibfield  {author} {\bibinfo {author} {\bibfnamefont {M.}~\bibnamefont
  {Asllani}}, \bibinfo {author} {\bibfnamefont {D.~M.}\ \bibnamefont
  {Busiello}}, \bibinfo {author} {\bibfnamefont {T.}~\bibnamefont {Carletti}},
  \bibinfo {author} {\bibfnamefont {D.}~\bibnamefont {Fanelli}}, \ and\
  \bibinfo {author} {\bibfnamefont {G.}~\bibnamefont {Planchon}},\ }\href@noop
  {} {\bibfield  {journal} {\bibinfo  {journal} {Scientific Reports}\ }\textbf
  {\bibinfo {volume} {5}},\ \bibinfo {pages} {1} (\bibinfo {year}
  {2015})}\BibitemShut {NoStop}%
\bibitem [{\citenamefont {Asllani}\ \emph {et~al.}(2016)\citenamefont
  {Asllani}, \citenamefont {Carletti},\ and\ \citenamefont
  {Fanelli}}]{Asllani2016}%
  \BibitemOpen
  \bibfield  {author} {\bibinfo {author} {\bibfnamefont {M.}~\bibnamefont
  {Asllani}}, \bibinfo {author} {\bibfnamefont {T.}~\bibnamefont {Carletti}}, \
  and\ \bibinfo {author} {\bibfnamefont {D.}~\bibnamefont {Fanelli}},\
  }\href@noop {} {\bibfield  {journal} {\bibinfo  {journal} {European Physical
  Journal B}\ }\textbf {\bibinfo {volume} {89}} (\bibinfo {year}
  {2016})}\BibitemShut {NoStop}%
\bibitem [{\citenamefont {Muolo}\ \emph {et~al.}(2019)\citenamefont {Muolo},
  \citenamefont {Asllani}, \citenamefont {Carletti}, \citenamefont {Fanelli},\
  and\ \citenamefont {Maini}}]{riccardo}%
  \BibitemOpen
  \bibfield  {author} {\bibinfo {author} {\bibfnamefont {R.}~\bibnamefont
  {Muolo}}, \bibinfo {author} {\bibfnamefont {M.}~\bibnamefont {Asllani}},
  \bibinfo {author} {\bibfnamefont {T.}~\bibnamefont {Carletti}}, \bibinfo
  {author} {\bibfnamefont {D.}~\bibnamefont {Fanelli}}, \ and\ \bibinfo
  {author} {\bibfnamefont {P.~K.}\ \bibnamefont {Maini}},\ }\href@noop {}
  {\bibfield  {journal} {\bibinfo  {journal} {Journal of Theoretical biology}\
  }\textbf {\bibinfo {volume} {480}} (\bibinfo {year} {2019})}\BibitemShut
  {NoStop}%
\bibitem [{\citenamefont {Asllani}\ \emph {et~al.}(2020)\citenamefont
  {Asllani}, \citenamefont {Carletti}, \citenamefont {Fanelli},\ and\
  \citenamefont {Maini}}]{AsllaniEPJB20}%
  \BibitemOpen
  \bibfield  {author} {\bibinfo {author} {\bibfnamefont {M.}~\bibnamefont
  {Asllani}}, \bibinfo {author} {\bibfnamefont {T.}~\bibnamefont {Carletti}},
  \bibinfo {author} {\bibfnamefont {D.}~\bibnamefont {Fanelli}}, \ and\
  \bibinfo {author} {\bibfnamefont {P.~K.}\ \bibnamefont {Maini}},\ }\href@noop
  {} {\bibfield  {journal} {\bibinfo  {journal} {European Physical Journal B}\
  }\textbf {\bibinfo {volume} {93}} (\bibinfo {year} {2020})}\BibitemShut
  {NoStop}%
\bibitem [{\citenamefont {H\"utt}\ \emph {et~al.}(2014)\citenamefont {H\"utt},
  \citenamefont {Kaiser},\ and\ \citenamefont {Hilgetag}}]{Hutt}%
  \BibitemOpen
  \bibfield  {author} {\bibinfo {author} {\bibfnamefont {M.-T.}\ \bibnamefont
  {H\"utt}}, \bibinfo {author} {\bibfnamefont {M.}~\bibnamefont {Kaiser}}, \
  and\ \bibinfo {author} {\bibfnamefont {C.~C.}\ \bibnamefont {Hilgetag}},\
  }\href@noop {} {\bibfield  {journal} {\bibinfo  {journal} {Philosophical
  Transactions of the Royal Society B}\ }\textbf {\bibinfo {volume} {369}}
  (\bibinfo {year} {2014})}\BibitemShut {NoStop}%
\bibitem [{\citenamefont {Watts}\ and\ \citenamefont
  {Strogatz}(1998)}]{Watts1998}%
  \BibitemOpen
  \bibfield  {author} {\bibinfo {author} {\bibfnamefont {D.}~\bibnamefont
  {Watts}}\ and\ \bibinfo {author} {\bibfnamefont {S.}~\bibnamefont
  {Strogatz}},\ }\href@noop {} {\bibfield  {journal} {\bibinfo  {journal}
  {Nature}\ }\textbf {\bibinfo {volume} {393}},\ \bibinfo {pages} {440}
  (\bibinfo {year} {1998})}\BibitemShut {NoStop}%
\bibitem [{\citenamefont {Meunier}\ \emph {et~al.}(2010)\citenamefont
  {Meunier}, \citenamefont {Lambiotte},\ and\ \citenamefont
  {Bullmore}}]{Meunier2010}%
  \BibitemOpen
  \bibfield  {author} {\bibinfo {author} {\bibfnamefont {D.}~\bibnamefont
  {Meunier}}, \bibinfo {author} {\bibfnamefont {R.}~\bibnamefont {Lambiotte}},
  \ and\ \bibinfo {author} {\bibfnamefont {E.~T.}\ \bibnamefont {Bullmore}},\
  }\href@noop {} {\bibfield  {journal} {\bibinfo  {journal} {Frontiers in
  Neuroscience}\ }\textbf {\bibinfo {volume} {4}},\ \bibinfo {pages} {1}
  (\bibinfo {year} {2010})}\BibitemShut {NoStop}%
\bibitem [{\citenamefont {Harriger}\ \emph {et~al.}(2012)\citenamefont
  {Harriger}, \citenamefont {Heuvel},\ and\ \citenamefont
  {Spoorns}}]{HarrigerLogan2012}%
  \BibitemOpen
  \bibfield  {author} {\bibinfo {author} {\bibfnamefont {L.}~\bibnamefont
  {Harriger}}, \bibinfo {author} {\bibfnamefont {M.~P. V.~D.}\ \bibnamefont
  {Heuvel}}, \ and\ \bibinfo {author} {\bibfnamefont {O.}~\bibnamefont
  {Spoorns}},\ }\href@noop {} {\bibfield  {journal} {\bibinfo  {journal} {PLoS
  ONE}\ } (\bibinfo {year} {2012})}\BibitemShut {NoStop}%
\bibitem [{\citenamefont {Hahn}\ \emph {et~al.}(2019)\citenamefont {Hahn},
  \citenamefont {Sporns}, \citenamefont {Watts},\ and\ \citenamefont
  {Swanson}}]{Hahn2019}%
  \BibitemOpen
  \bibfield  {author} {\bibinfo {author} {\bibfnamefont {J.~D.}\ \bibnamefont
  {Hahn}}, \bibinfo {author} {\bibfnamefont {O.}~\bibnamefont {Sporns}},
  \bibinfo {author} {\bibfnamefont {A.~G.}\ \bibnamefont {Watts}}, \ and\
  \bibinfo {author} {\bibfnamefont {L.~W.}\ \bibnamefont {Swanson}},\
  }\href@noop {} {\bibfield  {journal} {\bibinfo  {journal} {Proceedings of the
  National Academy of Sciences}\ }\textbf {\bibinfo {volume} {116}},\ \bibinfo
  {pages} {8018} (\bibinfo {year} {2019})}\BibitemShut {NoStop}%
\bibitem [{\citenamefont {Sporns}(2010)}]{SpornsBook}%
  \BibitemOpen
  \bibfield  {author} {\bibinfo {author} {\bibfnamefont {O.}~\bibnamefont
  {Sporns}},\ }\href@noop {} {\emph {\bibinfo {title} {{Networks of the
  Brain}}}}\ (\bibinfo  {publisher} {MIT Press},\ \bibinfo {year}
  {2010})\BibitemShut {NoStop}%
\bibitem [{\citenamefont {Simon}(1962)}]{Simon1962}%
  \BibitemOpen
  \bibfield  {author} {\bibinfo {author} {\bibfnamefont {H.~A.}\ \bibnamefont
  {Simon}},\ }\href@noop {} {\bibfield  {journal} {\bibinfo  {journal}
  {Proceedings of the American Philosophical Society}\ }\textbf {\bibinfo
  {volume} {106}},\ \bibinfo {pages} {467} (\bibinfo {year}
  {1962})}\BibitemShut {NoStop}%
\bibitem [{\citenamefont {Sporns}\ \emph {et~al.}(2007)\citenamefont {Sporns},
  \citenamefont {Honey},\ and\ \citenamefont {K{\"{o}}tter}}]{Sporns2007}%
  \BibitemOpen
  \bibfield  {author} {\bibinfo {author} {\bibfnamefont {O.}~\bibnamefont
  {Sporns}}, \bibinfo {author} {\bibfnamefont {C.~J.}\ \bibnamefont {Honey}}, \
  and\ \bibinfo {author} {\bibfnamefont {R.}~\bibnamefont {K{\"{o}}tter}},\
  }\href@noop {} {\bibfield  {journal} {\bibinfo  {journal} {PLoS ONE}\
  }\textbf {\bibinfo {volume} {2}},\ \bibinfo {pages} {1} (\bibinfo {year}
  {2007})}\BibitemShut {NoStop}%
\bibitem [{\citenamefont {Bullmore}\ and\ \citenamefont
  {Sporns}(2009)}]{Bullmore2009}%
  \BibitemOpen
  \bibfield  {author} {\bibinfo {author} {\bibfnamefont {E.}~\bibnamefont
  {Bullmore}}\ and\ \bibinfo {author} {\bibfnamefont {O.}~\bibnamefont
  {Sporns}},\ }\href@noop {} {\bibfield  {journal} {\bibinfo  {journal} {Nature
  reviews. Neuroscience}\ }\textbf {\bibinfo {volume} {10}},\ \bibinfo {pages}
  {186} (\bibinfo {year} {2009})}\BibitemShut {NoStop}%
\bibitem [{\citenamefont {Girvan}\ and\ \citenamefont
  {Newman}(2002)}]{NewmanPNAS}%
  \BibitemOpen
  \bibfield  {author} {\bibinfo {author} {\bibfnamefont {M.}~\bibnamefont
  {Girvan}}\ and\ \bibinfo {author} {\bibfnamefont {M.~E.}\ \bibnamefont
  {Newman}},\ }\href@noop {} {\bibfield  {journal} {\bibinfo  {journal} {PNAS}\
  }\textbf {\bibinfo {volume} {99}},\ \bibinfo {pages} {7821} (\bibinfo {year}
  {2002})}\BibitemShut {NoStop}%
\bibitem [{\citenamefont {Luo}\ \emph {et~al.}(2007)\citenamefont {Luo},
  \citenamefont {Yang}, \citenamefont {Chen}, \citenamefont {Chang},
  \citenamefont {Zhou},\ and\ \citenamefont {Scheuermann}}]{protein}%
  \BibitemOpen
  \bibfield  {author} {\bibinfo {author} {\bibfnamefont {F.}~\bibnamefont
  {Luo}}, \bibinfo {author} {\bibfnamefont {Y.}~\bibnamefont {Yang}}, \bibinfo
  {author} {\bibfnamefont {C.-F.}\ \bibnamefont {Chen}}, \bibinfo {author}
  {\bibfnamefont {R.}~\bibnamefont {Chang}}, \bibinfo {author} {\bibfnamefont
  {J.}~\bibnamefont {Zhou}}, \ and\ \bibinfo {author} {\bibfnamefont {R.~H.}\
  \bibnamefont {Scheuermann}},\ }\href@noop {} {\bibfield  {journal} {\bibinfo
  {journal} {Bioinformatics}\ }\textbf {\bibinfo {volume} {23}},\ \bibinfo
  {pages} {207} (\bibinfo {year} {2007})}\BibitemShut {NoStop}%
\bibitem [{\citenamefont {Jeong}\ \emph {et~al.}(2000)\citenamefont {Jeong},
  \citenamefont {Tombor}, \citenamefont {Albert}, \citenamefont {Oltvai},\ and\
  \citenamefont {Barab\'asi}}]{metabolic}%
  \BibitemOpen
  \bibfield  {author} {\bibinfo {author} {\bibfnamefont {H.}~\bibnamefont
  {Jeong}}, \bibinfo {author} {\bibfnamefont {B.}~\bibnamefont {Tombor}},
  \bibinfo {author} {\bibfnamefont {R.}~\bibnamefont {Albert}}, \bibinfo
  {author} {\bibfnamefont {Z.~N.}\ \bibnamefont {Oltvai}}, \ and\ \bibinfo
  {author} {\bibfnamefont {A.-L.}\ \bibnamefont {Barab\'asi}},\ }\href@noop {}
  {\bibfield  {journal} {\bibinfo  {journal} {Nature}\ }\textbf {\bibinfo
  {volume} {407}},\ \bibinfo {pages} {651} (\bibinfo {year}
  {2000})}\BibitemShut {NoStop}%
\bibitem [{\citenamefont {Redner}(1998)}]{Redner}%
  \BibitemOpen
  \bibfield  {author} {\bibinfo {author} {\bibfnamefont {S.}~\bibnamefont
  {Redner}},\ }\href@noop {} {\bibfield  {journal} {\bibinfo  {journal} {The
  European Physical Journal B -- Condensed Matter and Complex Systems}\
  }\textbf {\bibinfo {volume} {4}},\ \bibinfo {pages} {131} (\bibinfo {year}
  {1998})}\BibitemShut {NoStop}%
\bibitem [{Note1()}]{Note1}%
  \BibitemOpen
  \bibinfo {note} {The definition of the spectral gap depends on the way one
  defines the Laplacian matrix \cite {NewmanBook}. In our case the spectrum of
  the Laplacian is non positive.}\BibitemShut {Stop}%
\bibitem [{\citenamefont {Fortunato}(2010)}]{Fortunato}%
  \BibitemOpen
  \bibfield  {author} {\bibinfo {author} {\bibfnamefont {S.}~\bibnamefont
  {Fortunato}},\ }\href@noop {} {\bibfield  {journal} {\bibinfo  {journal}
  {Phys. Rep.}\ }\textbf {\bibinfo {volume} {486}},\ \bibinfo {pages} {75}
  (\bibinfo {year} {2010})}\BibitemShut {NoStop}%
\bibitem [{\citenamefont {Newman}\ and\ \citenamefont
  {Girvan}(2004)}]{Newman2004}%
  \BibitemOpen
  \bibfield  {author} {\bibinfo {author} {\bibfnamefont {M.~E.}\ \bibnamefont
  {Newman}}\ and\ \bibinfo {author} {\bibfnamefont {M.}~\bibnamefont
  {Girvan}},\ }\href@noop {} {\bibfield  {journal} {\bibinfo  {journal}
  {Physical Review E - Statistical, Nonlinear, and Soft Matter Physics}\
  }\textbf {\bibinfo {volume} {69}},\ \bibinfo {pages} {1} (\bibinfo {year}
  {2004})}\BibitemShut {NoStop}%
\bibitem [{\citenamefont {Newman}(2018)}]{NewmanBook}%
  \BibitemOpen
  \bibfield  {author} {\bibinfo {author} {\bibfnamefont {M.~E.~J.}\
  \bibnamefont {Newman}},\ }\href@noop {} {\emph {\bibinfo {title} {{Networks:
  An Introduction}}}},\ \bibinfo {edition} {2nd}\ ed.\ (\bibinfo  {publisher}
  {Oxford University Press},\ \bibinfo {year} {2018})\BibitemShut {NoStop}%
\bibitem [{\citenamefont {Vastano}\ \emph {et~al.}(1987)\citenamefont
  {Vastano}, \citenamefont {Pearson}, \citenamefont {Horsthemke},\ and\
  \citenamefont {Swinney}}]{Horsthemke1987}%
  \BibitemOpen
  \bibfield  {author} {\bibinfo {author} {\bibfnamefont {J.~A.}\ \bibnamefont
  {Vastano}}, \bibinfo {author} {\bibfnamefont {J.~E.}\ \bibnamefont
  {Pearson}}, \bibinfo {author} {\bibfnamefont {W.}~\bibnamefont {Horsthemke}},
  \ and\ \bibinfo {author} {\bibfnamefont {H.~L.}\ \bibnamefont {Swinney}},\
  }\href@noop {} {\bibfield  {journal} {\bibinfo  {journal} {Physics Letters
  A}\ }\textbf {\bibinfo {volume} {124}},\ \bibinfo {pages} {320} (\bibinfo
  {year} {1987})}\BibitemShut {NoStop}%
\bibitem [{\citenamefont {Pearson}\ and\ \citenamefont
  {Horsthemke}(1989)}]{Pearson}%
  \BibitemOpen
  \bibfield  {author} {\bibinfo {author} {\bibfnamefont {J.~E.}\ \bibnamefont
  {Pearson}}\ and\ \bibinfo {author} {\bibfnamefont {W.}~\bibnamefont
  {Horsthemke}},\ }\href@noop {} {\bibfield  {journal} {\bibinfo  {journal} {J.
  Chem. Phys.}\ }\textbf {\bibinfo {volume} {90}},\ \bibinfo {pages} {1588}
  (\bibinfo {year} {1989})}\BibitemShut {NoStop}%
\bibitem [{\citenamefont {V.~Castets}\ and\ \citenamefont
  {Kepper}(1990)}]{exp_Turing1}%
  \BibitemOpen
  \bibfield  {author} {\bibinfo {author} {\bibfnamefont {J.~B.}\ \bibnamefont
  {V.~Castets}, \bibfnamefont {E.~Dulos}}\ and\ \bibinfo {author}
  {\bibfnamefont {P.~D.}\ \bibnamefont {Kepper}},\ }\href@noop {} {\bibfield
  {journal} {\bibinfo  {journal} {Phys. Rev. Lett.}\ }\textbf {\bibinfo
  {volume} {64}},\ \bibinfo {pages} {2953} (\bibinfo {year}
  {1990})}\BibitemShut {NoStop}%
\bibitem [{\citenamefont {J.~Horváth}\ and\ \citenamefont
  {Kepper}(2009)}]{exp_Turing2}%
  \BibitemOpen
  \bibfield  {author} {\bibinfo {author} {\bibfnamefont {I.~S.}\ \bibnamefont
  {J.~Horváth}}\ and\ \bibinfo {author} {\bibfnamefont {P.~D.}\ \bibnamefont
  {Kepper}},\ }\href@noop {} {\bibfield  {journal} {\bibinfo  {journal}
  {Science}\ }\textbf {\bibinfo {volume} {324}},\ \bibinfo {pages} {772}
  (\bibinfo {year} {2009})}\BibitemShut {NoStop}%
\bibitem [{\citenamefont {Newman}(2006)}]{NewmanFiedler}%
  \BibitemOpen
  \bibfield  {author} {\bibinfo {author} {\bibfnamefont {M.~E.~J.}\
  \bibnamefont {Newman}},\ }\href@noop {} {\bibfield  {journal} {\bibinfo
  {journal} {Physical Review E}\ }\textbf {\bibinfo {volume} {74}},\ \bibinfo
  {pages} {036104} (\bibinfo {year} {2006})}\BibitemShut {NoStop}%
\bibitem [{\citenamefont {Reichardt}\ and\ \citenamefont
  {Bornholdt}(2006)}]{Reichardt2006}%
  \BibitemOpen
  \bibfield  {author} {\bibinfo {author} {\bibfnamefont {J.}~\bibnamefont
  {Reichardt}}\ and\ \bibinfo {author} {\bibfnamefont {S.}~\bibnamefont
  {Bornholdt}},\ }\href@noop {} {\bibfield  {journal} {\bibinfo  {journal}
  {Phys. Rev. E}\ }\textbf {\bibinfo {volume} {74}},\ \bibinfo {pages} {016110}
  (\bibinfo {year} {2006})}\BibitemShut {NoStop}%
\bibitem [{\citenamefont {Sporns}\ and\ \citenamefont
  {Zwi}(2004)}]{Sporns_Zwi}%
  \BibitemOpen
  \bibfield  {author} {\bibinfo {author} {\bibfnamefont {O.}~\bibnamefont
  {Sporns}}\ and\ \bibinfo {author} {\bibfnamefont {J.~D.}\ \bibnamefont
  {Zwi}},\ }\href@noop {} {\bibfield  {journal} {\bibinfo  {journal}
  {Neuroinformatics}\ }\textbf {\bibinfo {volume} {2}},\ \bibinfo {pages} {145}
  (\bibinfo {year} {2004})}\BibitemShut {NoStop}%
\bibitem [{\citenamefont {Albert}\ and\ \citenamefont
  {Barab\'asi}(2002)}]{albert_barabsi}%
  \BibitemOpen
  \bibfield  {author} {\bibinfo {author} {\bibfnamefont {R.}~\bibnamefont
  {Albert}}\ and\ \bibinfo {author} {\bibfnamefont {A.~L.~S.}\ \bibnamefont
  {Barab\'asi}},\ }\href@noop {} {\bibfield  {journal} {\bibinfo  {journal}
  {Rev. Mod. Phys}\ }\textbf {\bibinfo {volume} {74}},\ \bibinfo {pages} {47}
  (\bibinfo {year} {2002})}\BibitemShut {NoStop}%
\bibitem [{\citenamefont {Sporns}\ \emph {et~al.}(2004)\citenamefont {Sporns},
  \citenamefont {Chialvo}, \citenamefont {Kaiser},\ and\ \citenamefont
  {Hilgetag}}]{Sporns2004}%
  \BibitemOpen
  \bibfield  {author} {\bibinfo {author} {\bibfnamefont {O.}~\bibnamefont
  {Sporns}}, \bibinfo {author} {\bibfnamefont {D.~R.}\ \bibnamefont {Chialvo}},
  \bibinfo {author} {\bibfnamefont {M.}~\bibnamefont {Kaiser}}, \ and\ \bibinfo
  {author} {\bibfnamefont {C.~C.}\ \bibnamefont {Hilgetag}},\ }\href@noop {}
  {\bibfield  {journal} {\bibinfo  {journal} {Trends in Cognitive Sciences}\
  }\textbf {\bibinfo {volume} {8}},\ \bibinfo {pages} {418} (\bibinfo {year}
  {2004})}\BibitemShut {NoStop}%
\bibitem [{\citenamefont {FitzHugh}(1961)}]{FitzHugh1961}%
  \BibitemOpen
  \bibfield  {author} {\bibinfo {author} {\bibfnamefont {R.}~\bibnamefont
  {FitzHugh}},\ }\href@noop {} {\bibfield  {journal} {\bibinfo  {journal}
  {Biophysical Journal}\ }\textbf {\bibinfo {volume} {1}},\ \bibinfo {pages}
  {445} (\bibinfo {year} {1961})}\BibitemShut {NoStop}%
\bibitem [{\citenamefont {Nagumo}\ \emph {et~al.}(1962)\citenamefont {Nagumo},
  \citenamefont {Arimoto},\ and\ \citenamefont {Yoshizawa}}]{Nagumo1962}%
  \BibitemOpen
  \bibfield  {author} {\bibinfo {author} {\bibfnamefont {J.}~\bibnamefont
  {Nagumo}}, \bibinfo {author} {\bibfnamefont {S.}~\bibnamefont {Arimoto}}, \
  and\ \bibinfo {author} {\bibfnamefont {S.}~\bibnamefont {Yoshizawa}},\
  }\href@noop {} {\bibfield  {journal} {\bibinfo  {journal} {Proceedings of the
  IRE}\ }\textbf {\bibinfo {volume} {50}},\ \bibinfo {pages} {2061} (\bibinfo
  {year} {1962})}\BibitemShut {NoStop}%
\bibitem [{\citenamefont {Peixoto}(2013)}]{Peixoto}%
  \BibitemOpen
  \bibfield  {author} {\bibinfo {author} {\bibfnamefont {T.~P.}\ \bibnamefont
  {Peixoto}},\ }\href@noop {} {\bibfield  {journal} {\bibinfo  {journal} {Phys.
  Rev. Lett.}\ }\textbf {\bibinfo {volume} {111}},\ \bibinfo {pages} {098701}
  (\bibinfo {year} {2013})}\BibitemShut {NoStop}%
\bibitem [{\citenamefont {Bojan}(1991)}]{Bojan1991}%
  \BibitemOpen
  \bibfield  {author} {\bibinfo {author} {\bibfnamefont {M.}~\bibnamefont
  {Bojan}},\ }\href@noop {} {\bibfield  {journal} {\bibinfo  {journal} {Graph
  Theory, Combinatorics, and Applications}\ }\textbf {\bibinfo {volume} {2}},\
  \bibinfo {pages} {871} (\bibinfo {year} {1991})}\BibitemShut {NoStop}%
\bibitem [{Note2()}]{Note2}%
  \BibitemOpen
  \bibinfo {note} {BlueWe want to emphasise that regular networks (e.g., rings)
  have a large diameter, too, having this way a small spectral gap. However,
  our focus here is on random networks which, apart from the modular ones, are
  characterised by a small diameter.}\BibitemShut {Stop}%
\bibitem [{\citenamefont {Golub}\ and\ \citenamefont {van Loan}(1996)}]{Golub}%
  \BibitemOpen
  \bibfield  {author} {\bibinfo {author} {\bibfnamefont {G.~H.}\ \bibnamefont
  {Golub}}\ and\ \bibinfo {author} {\bibfnamefont {C.~F.}\ \bibnamefont {van
  Loan}},\ }\href@noop {} {\emph {\bibinfo {title} {{Matrix Computations}}}},\
  \bibinfo {edition} {3rd}\ ed.\ (\bibinfo  {publisher} {Johns Hopkins
  University Press},\ \bibinfo {year} {1996})\BibitemShut {NoStop}%
\bibitem [{Note3()}]{Note3}%
  \BibitemOpen
  \bibinfo {note} {Notice also that due to the algebraic connectivity, a
  network with a Laplacian spectral gap will always be modular.}\BibitemShut
  {Stop}%
\bibitem [{\citenamefont {Donetti}\ and\ \citenamefont {noz.}(2004)}]{Donetti}%
  \BibitemOpen
  \bibfield  {author} {\bibinfo {author} {\bibfnamefont {L.}~\bibnamefont
  {Donetti}}\ and\ \bibinfo {author} {\bibfnamefont {M.~A.~M.}\ \bibnamefont
  {noz.}},\ }\href@noop {} {\bibfield  {journal} {\bibinfo  {journal} {J. Stat.
  Mech.: Theor. Exp.}\ }\textbf {\bibinfo {volume} {2004}},\ \bibinfo {pages}
  {P10012} (\bibinfo {year} {2004})}\BibitemShut {NoStop}%
\bibitem [{\citenamefont {Andreotti}\ \emph {et~al.}(2018)\citenamefont
  {Andreotti}, \citenamefont {Remondini}, \citenamefont {Servizi},\ and\
  \citenamefont {Bazzani}}]{Andreotti}%
  \BibitemOpen
  \bibfield  {author} {\bibinfo {author} {\bibfnamefont {E.}~\bibnamefont
  {Andreotti}}, \bibinfo {author} {\bibfnamefont {D.}~\bibnamefont
  {Remondini}}, \bibinfo {author} {\bibfnamefont {G.}~\bibnamefont {Servizi}},
  \ and\ \bibinfo {author} {\bibfnamefont {A.}~\bibnamefont {Bazzani}},\
  }\href@noop {} {\bibfield  {journal} {\bibinfo  {journal} {Linear Algebra
  Appl.}\ }\textbf {\bibinfo {volume} {544}},\ \bibinfo {pages} {206} (\bibinfo
  {year} {2018})}\BibitemShut {NoStop}%
\bibitem [{\citenamefont {Smith}\ \emph {et~al.}(2018)\citenamefont {Smith},
  \citenamefont {Hein}, \citenamefont {Whitney}, \citenamefont {Fitzpatrick},\
  and\ \citenamefont {Kaschube}}]{Smith2018}%
  \BibitemOpen
  \bibfield  {author} {\bibinfo {author} {\bibfnamefont {G.~B.}\ \bibnamefont
  {Smith}}, \bibinfo {author} {\bibfnamefont {B.}~\bibnamefont {Hein}},
  \bibinfo {author} {\bibfnamefont {D.~E.}\ \bibnamefont {Whitney}}, \bibinfo
  {author} {\bibfnamefont {D.}~\bibnamefont {Fitzpatrick}}, \ and\ \bibinfo
  {author} {\bibfnamefont {M.}~\bibnamefont {Kaschube}},\ }\href@noop {}
  {\bibfield  {journal} {\bibinfo  {journal} {Nature Neuroscience}\ }\textbf
  {\bibinfo {volume} {21}},\ \bibinfo {pages} {1600–1608} (\bibinfo {year}
  {2018})}\BibitemShut {NoStop}%
\bibitem [{\citenamefont {Baniqued}\ \emph {et~al.}(2018)\citenamefont
  {Baniqued}, \citenamefont {Gallen}, \citenamefont {Voss}, \citenamefont
  {Burzynska}, \citenamefont {Wong}, \citenamefont {Cooke}, \citenamefont
  {Duffy}, \citenamefont {Fanning}, \citenamefont {Ehlers},\ and\ \citenamefont
  {Salerno}}]{Oldguys}%
  \BibitemOpen
  \bibfield  {author} {\bibinfo {author} {\bibfnamefont {P.~L.}\ \bibnamefont
  {Baniqued}}, \bibinfo {author} {\bibfnamefont {C.~L.}\ \bibnamefont
  {Gallen}}, \bibinfo {author} {\bibfnamefont {M.~W.}\ \bibnamefont {Voss}},
  \bibinfo {author} {\bibfnamefont {A.~Z.}\ \bibnamefont {Burzynska}}, \bibinfo
  {author} {\bibfnamefont {C.~N.}\ \bibnamefont {Wong}}, \bibinfo {author}
  {\bibfnamefont {G.~E.}\ \bibnamefont {Cooke}}, \bibinfo {author}
  {\bibfnamefont {K.}~\bibnamefont {Duffy}}, \bibinfo {author} {\bibfnamefont
  {J.}~\bibnamefont {Fanning}}, \bibinfo {author} {\bibfnamefont {D.~K.}\
  \bibnamefont {Ehlers}}, \ and\ \bibinfo {author} {\bibfnamefont {E.~A.}\
  \bibnamefont {Salerno}},\ }\href@noop {} {\bibfield  {journal} {\bibinfo
  {journal} {Frontiers in Aging Neuroscience}\ }\textbf {\bibinfo {volume} {9}}
  (\bibinfo {year} {2018})}\BibitemShut {NoStop}%
\bibitem [{Note4()}]{Note4}%
  \BibitemOpen
  \bibinfo {note} {Notice here that it may be, as in the case of Fig. \ref
  {fig:eigen_patt}, that different modules might share by chance the same level
  of components. However, this should not be understood as these entries
  belonging to the same module.}\BibitemShut {Stop}%
\bibitem [{\citenamefont {Schnabel}\ \emph {et~al.}(1973)\citenamefont
  {Schnabel}, \citenamefont {Bischoff}, \citenamefont {Hintze}, \citenamefont
  {Schulz}, \citenamefont {Hejnol}, \citenamefont {Meinhardt},\ and\
  \citenamefont {Hutter}}]{Fiedler}%
  \BibitemOpen
  \bibfield  {author} {\bibinfo {author} {\bibfnamefont {R.}~\bibnamefont
  {Schnabel}}, \bibinfo {author} {\bibfnamefont {M.}~\bibnamefont {Bischoff}},
  \bibinfo {author} {\bibfnamefont {A.}~\bibnamefont {Hintze}}, \bibinfo
  {author} {\bibfnamefont {A.-k.}\ \bibnamefont {Schulz}}, \bibinfo {author}
  {\bibfnamefont {A.}~\bibnamefont {Hejnol}}, \bibinfo {author} {\bibfnamefont
  {H.}~\bibnamefont {Meinhardt}}, \ and\ \bibinfo {author} {\bibfnamefont
  {H.}~\bibnamefont {Hutter}},\ }\href@noop {} {\bibfield  {journal} {\bibinfo
  {journal} {Czechoslovak Mathematical Journal}\ }\textbf {\bibinfo {volume}
  {23}},\ \bibinfo {pages} {298} (\bibinfo {year} {1973})}\BibitemShut
  {NoStop}%
\bibitem [{\citenamefont {Chung.}(1997)}]{Chung}%
  \BibitemOpen
  \bibfield  {author} {\bibinfo {author} {\bibfnamefont {F.}~\bibnamefont
  {Chung.}},\ }\href@noop {} {\emph {\bibinfo {title} {{Spectral Graph
  Theory}}}}\ (\bibinfo  {publisher} {Amer. Math. Soc.},\ \bibinfo {year}
  {1997})\BibitemShut {NoStop}%
\bibitem [{\citenamefont {Bumbarger}\ \emph {et~al.}(2013)\citenamefont
  {Bumbarger}, \citenamefont {Riebesell}, \citenamefont {Rödelsperger},\ and\
  \citenamefont {Sommer}}]{ppacificus}%
  \BibitemOpen
  \bibfield  {author} {\bibinfo {author} {\bibfnamefont {D.~J.}\ \bibnamefont
  {Bumbarger}}, \bibinfo {author} {\bibfnamefont {M.}~\bibnamefont
  {Riebesell}}, \bibinfo {author} {\bibfnamefont {C.}~\bibnamefont
  {Rödelsperger}}, \ and\ \bibinfo {author} {\bibfnamefont {R.~J.}\
  \bibnamefont {Sommer}},\ }\href@noop {} {\bibfield  {journal} {\bibinfo
  {journal} {Cell}\ }\textbf {\bibinfo {volume} {152}},\ \bibinfo {pages} {109
  } (\bibinfo {year} {2013})}\BibitemShut {NoStop}%
\bibitem [{\citenamefont {Sundaresan}\ \emph {et~al.}(2006)\citenamefont
  {Sundaresan}, \citenamefont {Fischhoff}, \citenamefont {Dushoff},\ and\
  \citenamefont {Rubenstein}}]{zebra}%
  \BibitemOpen
  \bibfield  {author} {\bibinfo {author} {\bibfnamefont {S.~R.}\ \bibnamefont
  {Sundaresan}}, \bibinfo {author} {\bibfnamefont {I.~R.}\ \bibnamefont
  {Fischhoff}}, \bibinfo {author} {\bibfnamefont {J.}~\bibnamefont {Dushoff}},
  \ and\ \bibinfo {author} {\bibfnamefont {D.~I.}\ \bibnamefont {Rubenstein}},\
  }\href@noop {} {\bibfield  {journal} {\bibinfo  {journal} {Oecologia}\
  }\textbf {\bibinfo {volume} {151}},\ \bibinfo {pages} {140–149} (\bibinfo
  {year} {2006})}\BibitemShut {NoStop}%
\bibitem [{\citenamefont {Sun}(2012)}]{Sun}%
  \BibitemOpen
  \bibfield  {author} {\bibinfo {author} {\bibfnamefont {G.}~\bibnamefont
  {Sun}},\ }\href@noop {} {\bibfield  {journal} {\bibinfo  {journal} {Nonlinear
  Dyn.}\ }\textbf {\bibinfo {volume} {69}},\ \bibinfo {pages} {1097} (\bibinfo
  {year} {2012})}\BibitemShut {NoStop}%
\bibitem [{\citenamefont {Sun}\ \emph {et~al.}(2016)\citenamefont {Sun},
  \citenamefont {Jusup}, \citenamefont {Jin}, \citenamefont {Wang},\ and\
  \citenamefont {Wang}}]{Sun_rev}%
  \BibitemOpen
  \bibfield  {author} {\bibinfo {author} {\bibfnamefont {G.}~\bibnamefont
  {Sun}}, \bibinfo {author} {\bibfnamefont {M.}~\bibnamefont {Jusup}}, \bibinfo
  {author} {\bibfnamefont {Z.}~\bibnamefont {Jin}}, \bibinfo {author}
  {\bibfnamefont {Y.}~\bibnamefont {Wang}}, \ and\ \bibinfo {author}
  {\bibfnamefont {Z.}~\bibnamefont {Wang}},\ }\href@noop {} {\bibfield
  {journal} {\bibinfo  {journal} {Phys. Life Rev.}\ }\textbf {\bibinfo {volume}
  {19}},\ \bibinfo {pages} {43} (\bibinfo {year} {2016})}\BibitemShut {NoStop}%
\bibitem [{kon(2016)}]{konectLink}%
  \BibitemOpen
  \href {http://konect.uni-koblenz.de/networks/tntp-ChicagoRegional} {\enquote
  {\bibinfo {title} {Chicago network dataset -- {KONECT}},}\ } (\bibinfo {year}
  {2016})\BibitemShut {NoStop}%
\bibitem [{\citenamefont {Eash}\ \emph {et~al.}(1983)\citenamefont {Eash},
  \citenamefont {Chon}, \citenamefont {Lee},\ and\ \citenamefont
  {Boyce}}]{konectArticle1}%
  \BibitemOpen
  \bibfield  {author} {\bibinfo {author} {\bibfnamefont {R.~W.}\ \bibnamefont
  {Eash}}, \bibinfo {author} {\bibfnamefont {K.~S.}\ \bibnamefont {Chon}},
  \bibinfo {author} {\bibfnamefont {Y.~J.}\ \bibnamefont {Lee}}, \ and\
  \bibinfo {author} {\bibfnamefont {D.~E.}\ \bibnamefont {Boyce}},\ }\href@noop
  {} {\bibfield  {journal} {\bibinfo  {journal} {Transportation Research
  Record}\ }\textbf {\bibinfo {volume} {994}},\ \bibinfo {pages} {30} (\bibinfo
  {year} {1983})}\BibitemShut {NoStop}%
\bibitem [{\citenamefont {Boyce}\ \emph {et~al.}(1985)\citenamefont {Boyce},
  \citenamefont {Chon}, \citenamefont {Ferris}, \citenamefont {Lee},
  \citenamefont {Lin},\ and\ \citenamefont {Eash}}]{konectArticle2}%
  \BibitemOpen
  \bibfield  {author} {\bibinfo {author} {\bibfnamefont {D.~E.}\ \bibnamefont
  {Boyce}}, \bibinfo {author} {\bibfnamefont {K.~S.}\ \bibnamefont {Chon}},
  \bibinfo {author} {\bibfnamefont {M.~E.}\ \bibnamefont {Ferris}}, \bibinfo
  {author} {\bibfnamefont {Y.~J.}\ \bibnamefont {Lee}}, \bibinfo {author}
  {\bibfnamefont {K.-T.}\ \bibnamefont {Lin}}, \ and\ \bibinfo {author}
  {\bibfnamefont {R.~W.}\ \bibnamefont {Eash}},\ }\href@noop {} {\bibfield
  {journal} {\bibinfo  {journal} {Chicago Area Transportation Study}\ ,\
  \bibinfo {pages} {xii + 169}} (\bibinfo {year} {1985})}\BibitemShut {NoStop}%
\bibitem [{\citenamefont {Kunegis}(2013)}]{konectProceedings}%
  \BibitemOpen
  \bibfield  {author} {\bibinfo {author} {\bibfnamefont {J.}~\bibnamefont
  {Kunegis}},\ }in\ \href
  {http://userpages.uni-koblenz.de/~kunegis/paper/kunegis-koblenz-network-collection.pdf}
  {\emph {\bibinfo {booktitle} {Proc. Int. Conf. on World Wide Web
  Companion}}}\ (\bibinfo {year} {2013})\ pp.\ \bibinfo {pages}
  {1343--1350}\BibitemShut {NoStop}%
\bibitem [{\citenamefont {Sporns}\ and\ \citenamefont {F.}(2016)}]{Sporns2016}%
  \BibitemOpen
  \bibfield  {author} {\bibinfo {author} {\bibfnamefont {O.}~\bibnamefont
  {Sporns}}\ and\ \bibinfo {author} {\bibfnamefont {B.~R.}\ \bibnamefont
  {F.}},\ }\href@noop {} {\bibfield  {journal} {\bibinfo  {journal} {Annu. Rev.
  Psychol.}\ }\textbf {\bibinfo {volume} {67}},\ \bibinfo {pages} {19.1}
  (\bibinfo {year} {2016})}\BibitemShut {NoStop}%
\bibitem [{\citenamefont {Asllani}\ \emph {et~al.}(2018)\citenamefont
  {Asllani}, \citenamefont {Expert},\ and\ \citenamefont
  {Carletti}}]{AsllaniPLOS}%
  \BibitemOpen
  \bibfield  {author} {\bibinfo {author} {\bibfnamefont {M.}~\bibnamefont
  {Asllani}}, \bibinfo {author} {\bibfnamefont {P.}~\bibnamefont {Expert}}, \
  and\ \bibinfo {author} {\bibfnamefont {T.}~\bibnamefont {Carletti}},\
  }\href@noop {} {\bibfield  {journal} {\bibinfo  {journal} {PLoS Computational
  Biology}\ ,\ \bibinfo {pages} {1}} (\bibinfo {year} {2018})}\BibitemShut
  {NoStop}%
\bibitem [{\citenamefont {Perc}(2005)}]{Perc2005}%
  \BibitemOpen
  \bibfield  {author} {\bibinfo {author} {\bibfnamefont {M.}~\bibnamefont
  {Perc}},\ }\href@noop {} {\bibfield  {journal} {\bibinfo  {journal} {New J.
  Phys.}\ }\textbf {\bibinfo {volume} {7}},\ \bibinfo {pages} {252} (\bibinfo
  {year} {2005})}\BibitemShut {NoStop}%
\bibitem [{\citenamefont {Ravasz}\ and\ \citenamefont
  {Barab\'asi}(2003)}]{hierarchical}%
  \BibitemOpen
  \bibfield  {author} {\bibinfo {author} {\bibfnamefont {E.~B.}\ \bibnamefont
  {Ravasz}}\ and\ \bibinfo {author} {\bibfnamefont {A.~L.~S.}\ \bibnamefont
  {Barab\'asi}},\ }\href@noop {} {\bibfield  {journal} {\bibinfo  {journal}
  {Physical Review E}\ }\textbf {\bibinfo {volume} {67}},\ \bibinfo {pages}
  {026112} (\bibinfo {year} {2003})}\BibitemShut {NoStop}%
\end{thebibliography}%

\end{document}